\newcommand{\be}{\begin{equation}}
\newcommand{\ee}{\end{equation}}
\newcommand{\bea}{\begin{eqnarray}}
\newcommand{\eea}{\end{eqnarray}}
\newcommand{\ba}[1]{\begin{array}{#1}}
\newcommand{\ea}{\end{array}}
\documentclass[preprint,12pt]{iopart}

\usepackage{epsfig}
\makeatletter
\expandafter\let\csname equation*\endcsname\relax
\expandafter\let\csname endequation*\endcsname\relax
\makeatother

\usepackage{amsmath}
\usepackage{amsmath}
\usepackage{amssymb}
\usepackage{mathrsfs}
\usepackage{color}
\begin{document}
\title{Hyperradiance, concurrence and photon blockade  in a pair of qubits inside a driven cavity}
\author{ Anushree Dey and Bimalendu Deb }
\address{School of Physical Sciences, Indian Association for the Cultivation of Science, Jadavpur, Kolkata 700032, India.}
\ead{msbd@iacs.res.in}
\begin{abstract}
We carry out a detailed theoretical study on the radiance, concurrence and photon blockade in the system of  a pair of qubits inside a high-quality cavity driven by a two-photon drive. Our results show that, in the weak driving regime, the system gives rise to strong hyperradiance along with finite concurrence or entanglement between the qubits. The entanglement persists even in the strong driving regime depending on the system parameters. We interpret these results in terms of two-qubit cavity-dressed states that are coupled  by the drive.  Additionally, we study the photon-photon correlations and field quadrature squeezing in the same set-up in the absence as well as in the presence of an intra-cavity Kerr-nonlinear medium. We  show  that the nonlinear medium leads to two-photon blockade.  Our results suggest that this system with nonlinearity may act as a quadrature-squeezed and hyperradiant two-photon source.

\end{abstract}
\maketitle
\section{Introduction}
Cooperative  effects of qubits interacting with a single-mode quantized field lead to the well-known phenomena of
Dicke superradiance \cite{PhysRev.93.99,PhysRevA.3.1735,PhysRevA.4.302,PhysRevLett.99.193602,PhysRevA.84.023805,PhysRevX.3.041001,PhysRevLett.76.2049,PhysRevA.7.831,PhysRevA.8.1440,hepp1973superradiant,hamner2014dicke,bohnet2012steady,PhysRevA.81.063827,bhatti2015superbunching} and subradiance \cite{PhysRevLett.115.243602,PhysRevLett.76.2049,PhysRevA.84.023805,PhysRevLett.116.083601,PhysRevLett.54.1917,PhysRevLett.108.123602}. Superradiance has  been observed in collective
emission from two-levels atoms \cite{baumann2010dicke,feng2015exploring,PhysRevLett.107.140402}, nuclei \cite{rohlsberger2010collective,rohlsberger2013cooperative} and pair of ions \cite{PhysRevLett.121.040503,PhysRevLett.76.2049}. In recent times, another cooperative  effect called hyperradiance (HR) \cite{li2022squeezed,xu2017hyperradiance,han2021hyperradiance,pleinert2017hyperradiance,li2021electromagnetic} wherein the emission rate
surpassses the superradiant rate of emission has come to the fore. Apart from  atomic or nuclear or any massive qubits, the massless photons in the context of cavity quantum electrodynamics (CQED) can also exhibit cooperative effects owing to some nonlinear processes. One such photonic cooperative effect is the photon blockade (PB) \cite{PhysRevLett.118.133604:Hampsen,PhysRevA.95.063842,PhysRevA.104.053718,PhysRevA.100.063817,PhysRevA.102.053710,PhysRevA.88.033836,PhysRevA.92.023838,PhysRevA.100.053857,PhysRevA.87.023809} which forbids the appearance of $(n+1)th$ photon in an $n$-photon blockade. It originates either from intrinsic nonlinearity associated with  the dressed-state ladder in the strong-coupling CQED \cite{PhysRevLett.118.133604:Hampsen,PhysRevA.95.063842,PhysRevA.100.063817,PhysRevA.102.053710,dey2023zero} or from the interaction of cavity photons with an intracavity nonlinear medium \cite{PhysRevA.104.053718,PhysRevA.100.053857,PhysRevA.87.023809,PhysRevA.92.023838,PhysRevX.10.021022,lin2020kerr,wang2018photon}.  Since a cavity field can mediate cooperative effects between  a  pair of atomic or ionic qubits inside a driven cavity, the question naturally arises if there is any interplay between the qubit-qubit and radiative cooperative effects.

 
 In this paper, we theoretically study a two-qubit
system coupled to single-mode cavity that is driven by a two-photon drive. A two-photon drive can be realized by pumping an intracavity crystal having second-order nonlinearity with an external laser of frequency tuned near twice the cavity frequency \cite{zhu2020squeezed}. This leads to an optical parametric amplification process and squeezing, whereby a single laser photon generates two cavity photons, thereby breaking the $U(1)$ symmetry of the system. The two-photon drive \cite{PhysRevLett.124.073602,PhysRevA.104.053718,PhysRevA.80.033807,zhu2020squeezed,wielinga1993quantum,minganti2016exact,zhao2018two,gevorgyan2013parametrically,PhysRevA.105.063704} involves processes where two photons are simultaneously created or annihilated. This does not conserve the photon number $\hat{n} = \langle a^{\dagger} a \rangle$. As a result, the Hamiltonian is no longer invariant under the $U(1)$ transformation  $a\rightarrow ae^{i\theta}$. The $U(1)$ symmetry corresponds to the conservation of particle number under this transformation.
 
 We quantify the radiance in terms of radiance witness $R$ as defined in Ref. \cite{pleinert2017hyperradiance}. The negativity of $R$ indicates subradiant regime implying that the radiation in the presence of two qubits is suppressed in comparison to that in the the presence of one qubit. The regime with $R=1$ is termed as superradiant regime. The regime with $R>1$ belongs to hyperradiant regime.  We calculate  concurrence as a measure of qubit-qubit entanglement. Concurrence is the most reliable  measure of two-qubit entanglement, and  has been used in the context of quantum cryptography  \cite{yin2020entanglement}, quantum teleportation \cite{llewellyn2020chip} and generation of quantum random number \cite{jacak2020quantum}.  We find that, in the weak driving regime where the driving rate is much smaller than the damping rate of the system, both
hyperradiance and concurrence appear in the system. We also demonstrate that, depending on the parameters, $R$ can reach a value as high as 50 which is substantially larger than the previously reported values \cite{li2022squeezed,xu2017hyperradiance,han2021hyperradiance,pleinert2017hyperradiance,li2021electromagnetic}. We interpret these results in terms of transitions between the two-qubit cavity-dressed states due to the drive and dissipation. Our analysis shows that the two-qubit entanglement results from the dressing of the two-qubit dressed states by the two-photon drive.  To explore the nonclassical properties of the field, we  calculate the Wigner distribution of the field \cite{PhysRev.40.749} which is found to be elliptical in the hyperradiant regime implying that the field is squeezed in quadrature. This quadrature squeezing is again confirmed by the klyshko's criterion \cite{klyshko1996observable} and the even-odd oscillations of the  photon number distributions. We further show that the photon statistics of the field is super-Poissonian in the parameter regime where both hyperradiance and quadrature squeezing are found. We  explore photon blockade (PB) in our system by introducing a Kerr nonlinear medium inside the cavity. PB is important for controlling photon statistics. In our model there exists no PB in the absence of a Kerr medium. The Kerr-nonlinear  resonator with a two-photon  drive has been studied in a variety of contexts \cite{wielinga1993quantum,minganti2016exact,zhao2018two,gevorgyan2013parametrically,PhysRevA.105.063704}.  The conventional  PB  occurs with a single atomic qubit or a pair of qubits inside a cavity in the strong coupling CQED regime \cite{PhysRevLett.118.133604:Hampsen,PhysRevA.95.063842}. We find that, in the parameter regime of hyperradiance  and quadrature squeezing, the Kerr nonlinearity leads to the two-photon blockade. Thus our proposed system with nonlinearity may act as a quadrature-squeezed and hyperradiant two-photon source with two-qubit entanglement in some selective parameter regime.

The rest of this paper is organized as follows: In Sec. \ref{model} we present our model and methods. In Sec.\ref{discussion} we present and  discuss our results. Finally, we conclude in Sec. \ref{conclusions}.

\section{The Model and the formulation of the problem}\label{model}
We consider a single-mode cavity with frequency $\omega_c$ interacting with a pair of identical qubits with transition frequency
$\omega_a$. The cavity is driven by a two-photon field
\begin{equation}
           H_d = \eta \left ({\hat{a}^{\dagger2}} e^{-2i\omega_dt}+{\hat{a}}^2 e^{2i\omega_dt} \right) 
\end{equation}
where $\hat{a} (\hat{a^{\dagger}})$ is the annihilation (creation) operator of the cavity field and $\eta$ is the driving strength. In cavity QED systems, a two-photon drive is typically implemented  by using a second order nonlinear medium (e.g., a $\chi^{(2)}$ crystal) inside the cavity and driving the cavity with an external field  twice the cavity's resonant frequency. The nonlinear medium inside the driven cavity enables parametric processes such as second-harmonic generation or two-photon down-conversion. A Kerr medium with a two-photon drive is a powerful platform in quantum optics for generating and stabilizing non-classical states of light \cite{puri2017engineering,PhysRevA.98.062313}. The Kerr nonlinearity introduces an intensity-dependent phase shift, while the two-photon drive enables the creation of quantum superpositions such as Schrödinger cat states \cite{he2023fast}. This combination is particularly important for quantum information processing \cite{kirchmair2013observation}, where it allows for robust encoding of qubits in bosonic modes and supports autonomous error correction. It is to be noted here that the similar model has been used previously both theoretically \cite{PhysRevA.105.013717,PhysRevA.80.033807,PhysRevLett.124.073602,faghihi2012dynamics,abdel2024dynamical,almalki2024interaction,ateto2015atom,PhysRevA.105.063704,PhysRevA.104.053718} and experimentally \cite{kirchmair2013observation}. Also, the similar model has been engineered using optical parametric oscillator (OPO). For example,  Refs. \cite{PhysRevA.105.013717, PhysRevA.80.033807,PhysRevLett.124.073602,PhysRevA.104.053718} have reported CQED studies with an intra-cavity Kerr nonlinear medium.
Our model system is schematically shown in Fig.1 where a cavity containing a second order ($\chi^2$) nonlinear medium and a pair of qubits is driven by a coherent drive. Additionally, the cavity may or may not contain a Kerr nonlinear medium. The coherent drive pumps the $\chi^2$ medium that produces the desirable two-photon drive.

In the frame rotating with the driving frequency $2\omega_d$, the Hamiltonian of the system takes the form 
\begin{equation}
\begin{split}
    H=\Delta_c {\hat a}^\dagger\hat{a}+\Delta_a\sum_{i=1,2}{\hat{\sigma_i}}^{\dagger}{\hat{\sigma_i}}+(g_i{\hat a}^\dagger{\hat{\sigma_i}}+\rm{h.c})+
    \chi {\hat{a}}^\dagger {\hat{a}}^\dagger \hat{a}\hat{a}+\eta({\hat{a}^{\dagger2}}+\hat{a}^2)
\end{split}
\end{equation}
where $g_i=g \cos(2\pi z_i/\lambda_c)$ with $z_i$ being the position of the $i$-th atom and $\lambda_c$ is the wavelength of the cavity field. Here $\Delta_a=\omega_a-\omega_d$ and $\Delta_c=\omega_c-\omega_d$. The dynamics of the system is governed by the master equation
\bea
\frac{\partial{\hat\rho}}{\partial t} = - \frac{i}{\hbar}\left[H,\hat\rho\right] + \frac{\kappa}{2}\mathcal{L}_{a}  + \sum_{i=1,2} \frac{\gamma_i}{2}\mathcal{L}_{\sigma_i}
 \label{eq1}
\eea
where $\rho$ is the density matrix of the system. $\cal L_{\rm{x}}$ is  a Liouville super operator where $x$ stands for any of the two operators $a,\sigma_j$ ($j=1,2$). Here the Lindbladian superoperators 
\begin{align}
\mathcal{L}_{a}&=\left[2a\rho a^\dagger- \left \{ a^\dagger a, \rho \right \}\right]~, \\
\mathcal{L}_{\sigma_j}&= \left[ 2\sigma_j\rho\sigma_j^{\dagger} - \left \{ \sigma_j^{\dagger} \sigma_j, \rho \right \} \right ]~, \nonumber 
 \end{align}
 describe the cavity and atomic damping, respectively with $\kappa$ and $\gamma$ being the cavity and atomic decay rates. We solve this master equation numerically \cite{johansson2012qutip} and calculate steady-state properties.
 \begin{figure}
 \centering
   \includegraphics[width=5in]{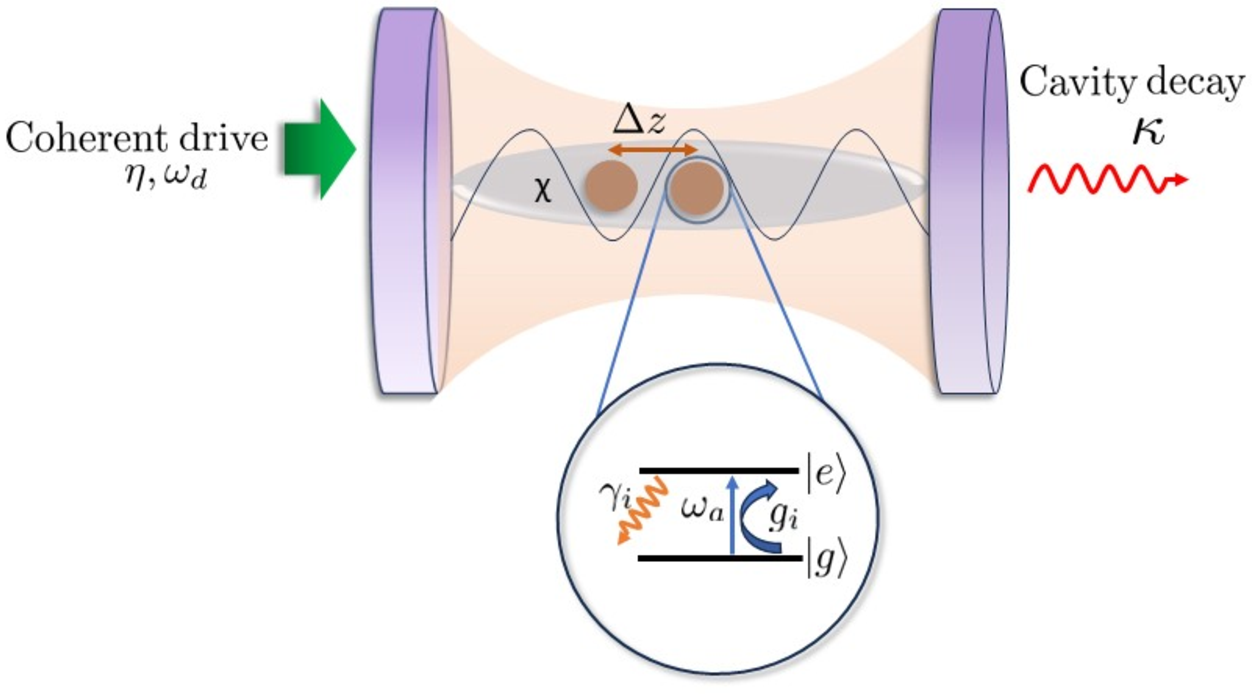}
    \caption{Schematic diagram of our system: a pair of identical qubits inside a cavity that may contain a Kerr nonlinear medium. Here, we have shown a pair of two-level atom as qubits. The cavity is driven by a two-photon drive with strength $\eta$ and frequency $\omega_d$. Atoms have the transition frequency $\omega_a$. The $i$-th atom is coupled to the cavity with strength $g_i$ and $\gamma_i$ is its decay rate. The decay rate of the cavity is $\kappa$.}
    \label{schematic}
\end{figure}

\begin{figure}
 \centering
 \vspace{.25in}
 \includegraphics[width=0.9\linewidth]{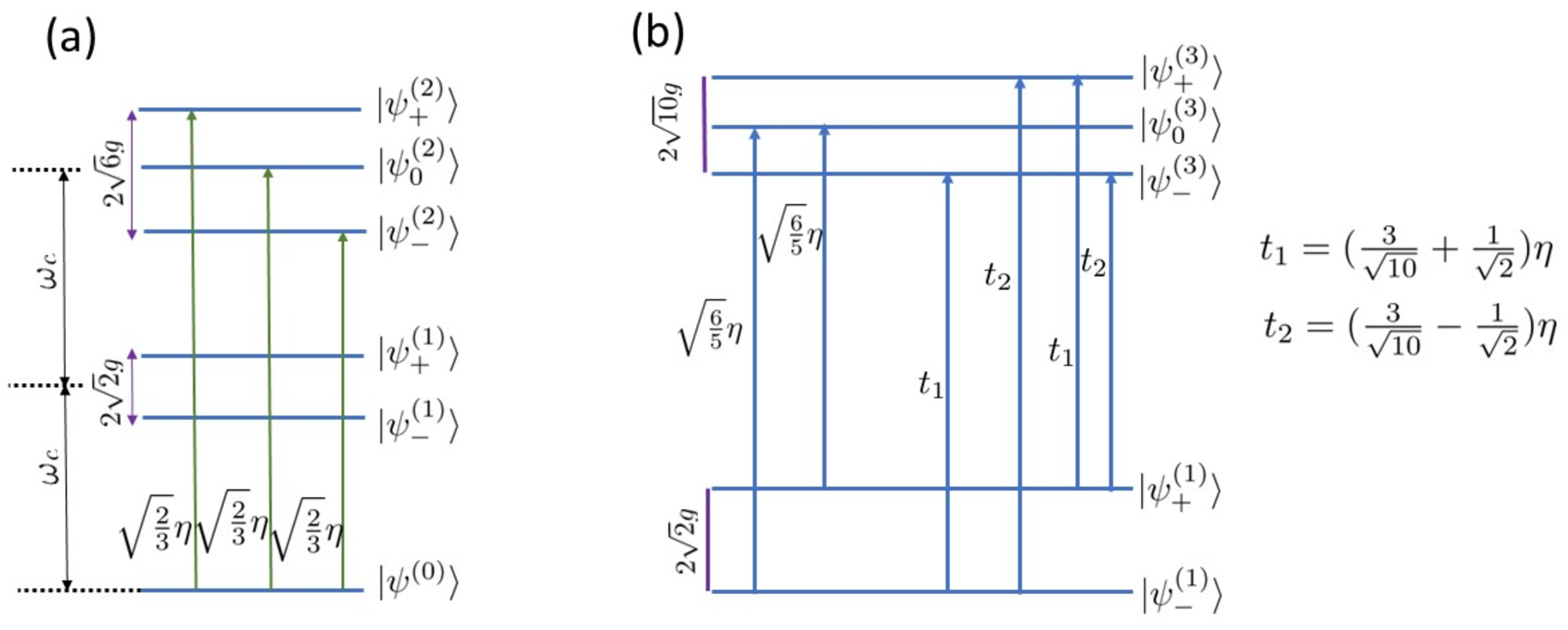}\\
 \caption{(a) Dressed-state level diagram of zero- and two-photons sectors for the coupled two-qubit-cavity system with three possible two-photon transitions. All these three transitions have the same matrix element $\sqrt{\frac{2}{3}}\eta$ with $\eta$ being the pump field Rabi frequency. (b) Dressed-state level diagram with transitions between one- and three-photon sectors.}
 \label{dressed_state}
 \end{figure}

In the absence of the drive and decay processes, one can develop a two-qubit dressed-state structure that can help to gain insight into the underlying physical processes in the system. Throughout our calculations, we have assumed that qubit-cavity coupling for both the atoms are same $g_1=g_2=g$. We have defined the collective two-qubit basis states as $|gg\rangle$,$|\pm\rangle=\frac{|ge>\pm|eg\rangle}{\sqrt{2}}$ and $|ee\rangle$ to describe the dynamics in the dressed-state picture. In the single-photon space the eigenvalues and eigenvectors (dressed states) are obtained as $|\psi_0^{(1)}\rangle=|-,0\rangle$ with $E_{dr}^{10}=0$ and $|\psi^{(1)}_{\pm}\rangle=\frac{|gg,1\rangle\pm|+,0\rangle}{\sqrt{2}}$ with $E_{dr}^{1\pm}=\pm\sqrt{2}g$. In the two-photon space, the  eigenvectors are $|\psi^{(2)}_{0}\rangle=-\frac{1}{\sqrt{3}}|gg,2\rangle+\frac{\sqrt{2}}{\sqrt{3}}|ee,0\rangle$ and $|\phi^{(2)}_{0}\rangle=|-,1\rangle$ with corresponding eigenvalue $E^{20}=0$. The another eigenvector is $\psi_{\pm}^{(2)}=\frac{1}{\sqrt{3}} |gg,2\rangle\pm \frac{1}{\sqrt{2}}|+,1\rangle+\frac{1}{\sqrt{6}}|ee,0\rangle$ with corresponding eigenvalue $E_{dr}^{2\pm}=\sqrt{6}g$. 
Due to the symmetric coupling ($g_1=g_2=g$) the antisymmetric Dicke state $|-,n\rangle$ is decoupled in the $n$ -photon space ($n=1,2$). In order to know the allowed transitions between the dressed states in the presence of a two-photon drive, we calculate the transition dipole matrix elements between the  dressed states as shown in Fig. \ref{dressed_state}.

 In the two-photon space the Hamiltonian in the basis $|gg,2\rangle,|\pm,1\rangle$ and $|ee,0\rangle$, taking $\omega_c=\omega_a$, is given by,
\begin{eqnarray}
H = \begin{bmatrix}
2\omega_c + 2\chi & 2g & 0 & 0 \\
2g & 2\omega_c & 0 & \sqrt{2}g \\
0 & 0 & 2\omega_c & 0 \\
0 & \sqrt{2}g & 0 & 2\omega_c
\end{bmatrix}.
\end{eqnarray}
The dressed energies and dressed states are obtained by diagonalizing this Hamiltonian. 

Figure. \ref{dressed_state} shows the two-qubit dressed-state level diagram and possible two-photon transitions between low-lying dressed levels. The rate equation of the density matrix elements between two dressed states \cite{cohen1998atom} can be obtained from master equation as follows

\begin{multline}
\langle \psi^{(2)}_0|\dot{\rho}|\psi_0^{(0)}\rangle =
-\left( \frac{2i\Delta}{\bar{h}} + \frac{\kappa}{3} + \frac{2\gamma}{3} \right) 
\langle \psi^{(2)}_0|\rho|\psi_0^{(0)}\rangle \\
+\frac{i}{\bar{h}} \sqrt{\frac{2}{3}}\,\eta 
\left( 
  \sum_{i=\pm}\langle \psi^{(2)}_0|\rho|\psi_{i}^{(2)}\rangle  
  - \langle \psi^{(2)}_0|\rho|\psi_0^{(2)}\rangle 
  + \langle \psi_0^{(0)}|\rho|\psi_0^{(0)}\rangle 
\right) \\
+ \sum_{i=\pm} \left( -\frac{\kappa}{2\sqrt{15}} - \frac{\gamma}{2\sqrt{3}} \right) 
  \langle \psi_{i}^{(3)}|\rho|\psi_{i}^{(1)}\rangle \\
+ \sum_{\substack{i,j=\pm \\ i\neq j}} \left( -\frac{\kappa}{2\sqrt{15}} + \frac{\gamma}{2\sqrt{3}} \right) 
  \langle \psi_{i}^{(3)}|\rho|\psi_{j}^{(1)}\rangle \\
+ \frac{2\kappa}{\sqrt{5}} 
 \sum_{i=\pm}
  \langle \psi_0^{(3)}|\rho|\psi_{i}^{(1)}\rangle \\
+ \left( \frac{\kappa}{3} - \frac{\gamma}{3} \right) \langle \psi_{+}^{(2)}|\rho|\psi_0^{(0)}\rangle
+ \left( \frac{\kappa}{3} + \frac{\gamma}{3} \right) \langle \psi_{-}^{(2)}|\rho|\psi_0^{(0)}\rangle
\label{rate eq.}
\end{multline}
The rate equation for $\langle \psi^{(2)}_{\pm}|\dot{\rho}|\psi_0^{(0)}\rangle $ is similar as above but $\mid \psi^{(2)}_{0} \rangle$ should be replaced by $\mid \psi^{(2)}_{\pm} \rangle$. We will employ these rate equations to explain our results.

 \begin{table}[!h]
     \centering
     \begin{tabular}{c|c|c}
         \hline
 Photon sector & Dressed energies ($\hbar \omega^{ij}_{dr}$) & Dressed states ($\psi_j^{i}$) \\ [0.8ex] 
 \hline\hline
 0 &$0$ & $|\psi_0^{(0)}\rangle=|gg,0\rangle$\\
 \hline 
 1 &$\sqrt{2}g$ & $|\psi_+^{(1)}\rangle=\frac{|gg,1\rangle+|+,0\rangle}{\sqrt{2}}$\\
  \hline 
 1 &$-\sqrt{2}g$ & $|\psi_-^{(1)}\rangle=\frac{|gg,1\rangle-|+,0\rangle}{\sqrt{2}}$\\
 \hline 
 2 &$\sqrt{6}g$ & $|\psi_+^{(2)}\rangle={\frac{1}{\sqrt{3}}|gg,2\rangle+\frac{1}{\sqrt{2}}|+,1\rangle+\frac{1}{\sqrt{6}}|ee,0\rangle}$\\
  \hline 
 2 &$-\sqrt{6}g$ & $|\psi_-^{(2)}\rangle={\frac{1}{\sqrt{3}}|gg,2\rangle-\frac{1}{\sqrt{2}}|+,1\rangle+\frac{1}{\sqrt{6}}|ee,0\rangle}$\\
 \hline 
  2 &$0$ & $|\phi_0^{(2)}\rangle={|-,1\rangle}$\\
 \hline 
  2 &$0$ & $|\psi_0^{(2)}\rangle=-{\frac{1}{\sqrt{3}}|gg,2\rangle}+\sqrt{\frac{2}{3}}|ee,0\rangle$\\
 \hline
 3 &$\sqrt{10}g$ & $\psi_{+}^{(3)}=\frac{\sqrt{3}}{\sqrt{10}}|gg,3\rangle+\frac{1}{\sqrt{2}}|+,2\rangle+\frac{1}{\sqrt{5}}|ee,1\rangle$\\
  \hline 
 3 &$-\sqrt{10}g$ & $\psi_{-}^{(3)}=\sqrt\frac{{3}}{{10}}|gg,3\rangle-\frac{1}{\sqrt{2}}|+,2\rangle+\frac{1}{\sqrt{5}}|ee,1\rangle$\\
 \hline 
  3 &$0$ & $\phi_0^{(3)}=|-,2\rangle$\\
   \hline 
  3 &$0$ & $\psi_0^{(3)}=-\sqrt\frac{2}{5}|gg,3\rangle+\frac{3}{\sqrt{15}}|ee,1\rangle$\\
  \hline
     \end{tabular}
    \caption{Dressed state energies and dressed states for different photon sectors.} 
     \label{tab:my_label}
 \end{table}

 \subsection{Radiance and concurrence}

 One can measure the radiant nature of the two-qubit system based on the correlations between the two qubits using the quantity called radiance witness $R$  \cite{pleinert2017hyperradiance,li2022squeezed} defined by
 \begin{equation}
    R=\frac{\langle a^{\dagger} a\rangle_2-2 \langle a^{\dagger} a\rangle_{1}}{2 \langle a^{\dagger} a\rangle_{1}}
\end{equation}
This involves the intracavity average photon number subject to the existence of one or two qubits inside the cavity. $\langle a^{\dagger} a\rangle_2$ denotes the average photon number when both the qubits are present in the cavity. The average photon number when only $i$th single qubit is present in the cavity is denoted by $\langle a^{\dagger} a\rangle_{1}$. $R=0$ implies that the radiation is emitted from uncorrelated qubits, whereas any finite value of $R$ indicates the existance of a correlation between the qubits. The negativity of $R$ implies that the radiation is suppressed compared to that due to two qubits, corresponding to the so-called subradiance regime, while $R > 0$
reveals that the radiation is enhanced due to the presence of both qubits. The value of $R=1$ specifically implies that the radiation exhibits scaling proportional to the square of the number of qubits as a characteristic feature of superradiance. When qubits are confined within a cavity, they experience a backaction from the cavity field. This modifies their collective radiative behavior, opening up an intriguing new possibility of hyperradiance regime characterised by $R>1$.
We numerically study the properties of radiance solving the master equation (\ref{eq1}). Furthermore, to gain some analytical insight about hyperradiance, we address the problem by wave function approach with a complex Hamiltonian in the weak driving regime as discussed in Appendix \ref{analytic}. In this approach, the mean photon numbers for single- and two-qubit-cavity system in steady-state are given by
\begin{eqnarray}
\langle {\hat{a}}^\dagger a\rangle_{1}&=&|C_{g,2}|^2+2|C_{e,1}|^2\nonumber\\
\langle {\hat{a}}^\dagger a\rangle_{2}&=&|C_{gg,1}|^2+2|C_{gg,2}|^2,
\end{eqnarray}
respectively, where $C_{gg,n}, C_{g,2}, C_{e,1}$ are the probability amplitudes of the states $|gg,n\rangle, |g,2\rangle, |e,1\rangle$ as defined in Appendix \ref{analytic}. We thus obtain a perturbative expression of $R$ given in the appendix.

To explore qubit-qubit entanglement we make use of the concurrence \cite{PhysRevLett.80.2245} defined by 
\begin{equation}
    C(\rho_q) = [\rm{max}\{{0,{\lambda_1}-{\lambda_2}-{\lambda_3}-{\lambda_4}}\}]
\end{equation} 
where ${\lambda_1,\lambda_2,\lambda_3,\lambda_4}$ are the eigenvalues of the matrix $R=\sqrt{\sqrt{\rho_q}\Tilde{\rho_q}\sqrt{\rho_q}}$ in decreasing order (the operator $\Tilde{\rho_q}= (\sigma_y \otimes \sigma_y) \rho_q^* (\sigma_y \otimes \sigma_y)$ , $\sigma_y$ being the Pauli matrix). Here, $\rho_q$ is the reduced density matrix of the two qubits. $C$ ranges from zero for unentangled states to one for maximally entangled states.
 We explore two-qubit entanglement in terms of concurrence and its possible connection with the radiance properties of the field.

 \subsection{$n$ photon blockade ($n$PB) and photon statistics } 
 
Apart from radiance-concurrence connection, it is also worth investigating any possible relation between radiance and $n$-photon blockade. By examining the steady-state photon-number distribution $P_n=\langle |n\rangle \langle n|\rangle=\rm {Tr}[\rho |n\rangle\langle n|]$ and the equal-time $n$th-order correlation function $g^{(n)}(0)=\frac{\langle{a^{\dagger}}^n a^n\rangle}{\langle a^{\dagger}a\rangle^n}$, it is possible to understand the physical origin of the $n$PB phenomena. Hamsen $\it et$$\it al.$ \cite{PhysRevLett.118.133604:Hampsen} measured the $n$PB effect and suggested two standards. The primary standard is based on a comparative analysis between the photon-number distribution of the system under study and the Poisson distribution. Specifically, the criterion for the occurrence of $n$PB is given by
\begin{equation}
    P_n\ge\Pi_n,P_{m>n}<\Pi_{m>n}
    \label{eq.npb_1}
\end{equation}
where $\Pi_n$ is the Poisson distribution given by
$\Pi_n=\langle n\rangle^n \exp(-\langle n\rangle)/n!$. $\langle n\rangle$ is the average number of photons given by $\langle n\rangle=\rm{Tr}[a^{\dagger}a\rho]$. According to equation (\ref{eq.npb_1}), the $n$PB effect pertains to an increase of the likelihood of $n$ photons while suppressing the probabilities of the photon numbers greater than $n$. The other criterion is based on $g^{(n)}(0)$, an equal-time nth-order correlation function. The $n$PB effect must meet the following requirements
\begin{equation}
    g^{(n)}(0)\ge1,g^{(n+1)}(0)<1
\end{equation}
meaning that the field has $(n + 1)$th order sub-Poissonian photon statistics and the $n$th order poissonian or super-Poissonian photon statistics. In particular, the 2PB effect satisfies the correlation functions $g^{(2)}(0)\ge 1$ and $g^{(3)}(0)<1$. The 1PB effect can be characterized by $g^{(2)}(0)<<1$. We have calculated $g^{(2)}(0)$ using the numerical solution of the master equation (\ref{eq1}) as well as the perturbative analytical solution as obtained in Appendix- \ref{analytic} through complex hamiltonian approach. Following the later procedure the equal-time second-order correlation function can be expressed as
\begin{equation}
g^{(2)}(0)=\frac{2|C_{gg,2}|^2}{(2|C_{gg,2}|^2+|C_{+,1}|^2)^2}
\label{g20_C}
\end{equation}
where $C_{gg,2}$ and $C_{+,1}$ are the probability amplitude of the states $|gg,2\rangle$ and $|+,1\rangle$ as defined in Appendix \ref{analytic}.
 In order to seek possible connection between $n$-photon blockade and the nonclassical nature of the cavity field, we calculate three quantities, namely squeezing parameter $S_{\theta}$, Klyshko's number $K_n$ and radiance witness $R$.
 
To describe the squeezing character of the cavity field,  we calculate the squeezing parameter 
\begin{equation}
    S_{\theta}=(\langle X_{\theta}^2\rangle-\langle X_{\theta}\rangle^2)-\frac{1}{2}
\end{equation}
where $X_{\theta}=\frac{(ae^{-i\theta}+a^{\dagger}e^{i\theta})}{\sqrt{2}}$ is the linear combination of the hermitian quadrature operators $X=\frac{(a+a^{\dagger})}{\sqrt{2}}$ and $Y=\frac{i(a^{\dagger}-a)}{\sqrt{2}}$.
Any negative value of $S_{\theta}$ implies squeezing nature of the field characterized by narrowing of the quadrature distribution compared to that of a coherent state.

To verify the nonclassicality of the photon number distribution, we use the Klyshko's criterion \cite{klyshko1996observable}
\begin{equation}
    K_n=\frac{(n+1)P_{n-1}P_{n+1}}{nP_n^2}
\end{equation}
where $P_n$ stands for the photon number distribution. The state is nonclassical when any $K_n$ value is below one. This introduces an additional criterion for determining the nonclassical nature of the field apart from $S_{\theta}$. In squeezed states, photons are predominantly distributed among even-numbered states within the Fock space, leading to an absence of odd-numbered photon states. The photon number distribution for squeezed light exhibits oscillations between even and odd numbers. Consequently, the Klyshko’s criterion incorporating odd-even oscillation somewhat mirrors the field's squeezing characteristics.


\section{Results and discussions}\label{discussion}

At the outset,  we set $\Delta_a=\Delta_c=\Delta$, that is, the frequency of the drive is equally tuned  from atomic and cavity frequencies, and the cavity frequency is on resonance with the atomic transition frequency. The unit of frequency is chosen to be the qubit decay constant $\gamma$, so all frequency quantities are assumed to be scaled by $\gamma$, unless otherwise stated. 

We first present the results in the absence of Kerr nonlinearity and then discuss the effects of this nonlinearity ($\chi$). 

\subsection{Results in the absence of a Kerr medium}\label{chi0}

\begin{figure}
 \centering
 \vspace{.25in}
 \includegraphics[width=0.44\linewidth]{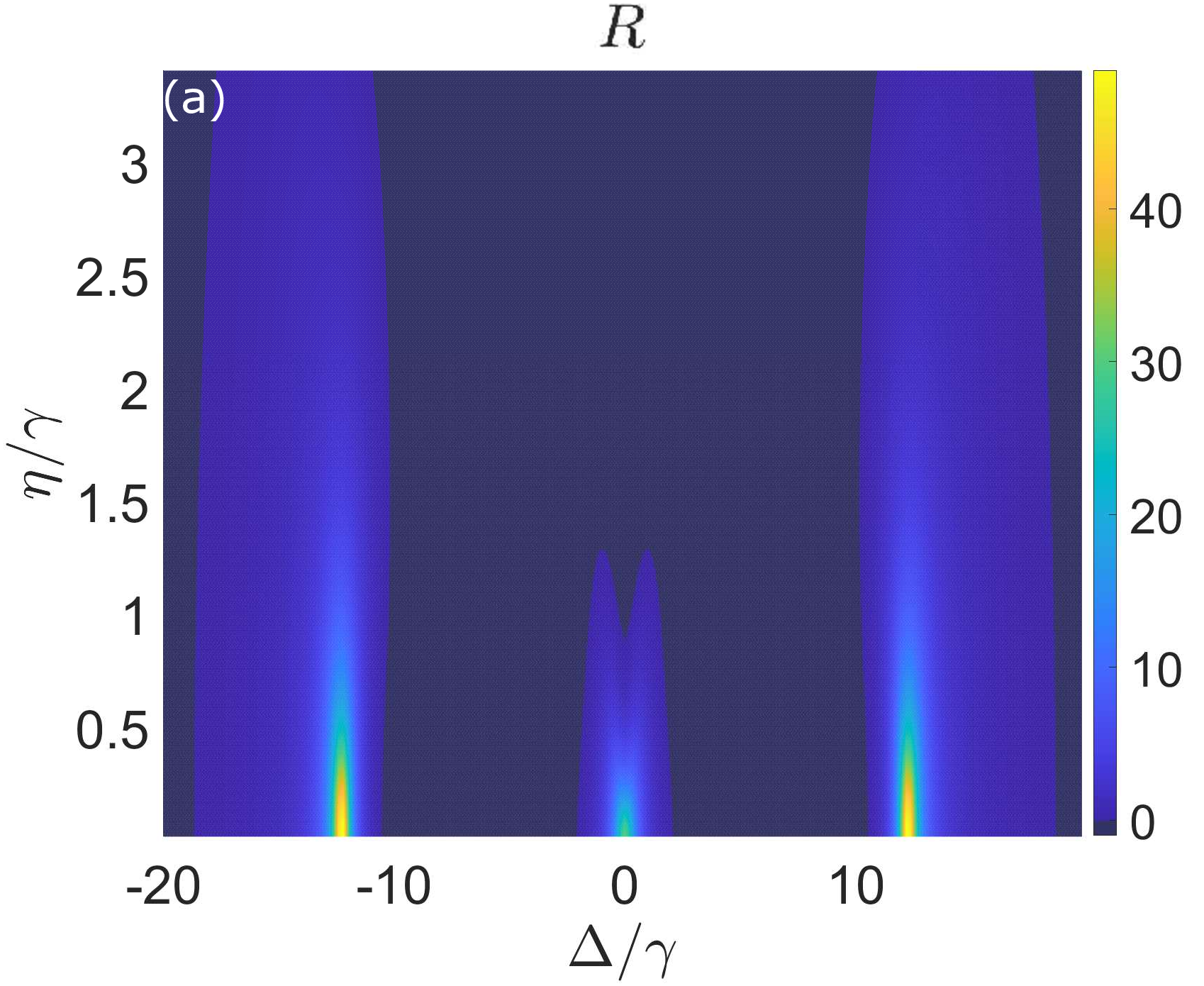}
 \includegraphics[width=0.45\linewidth]{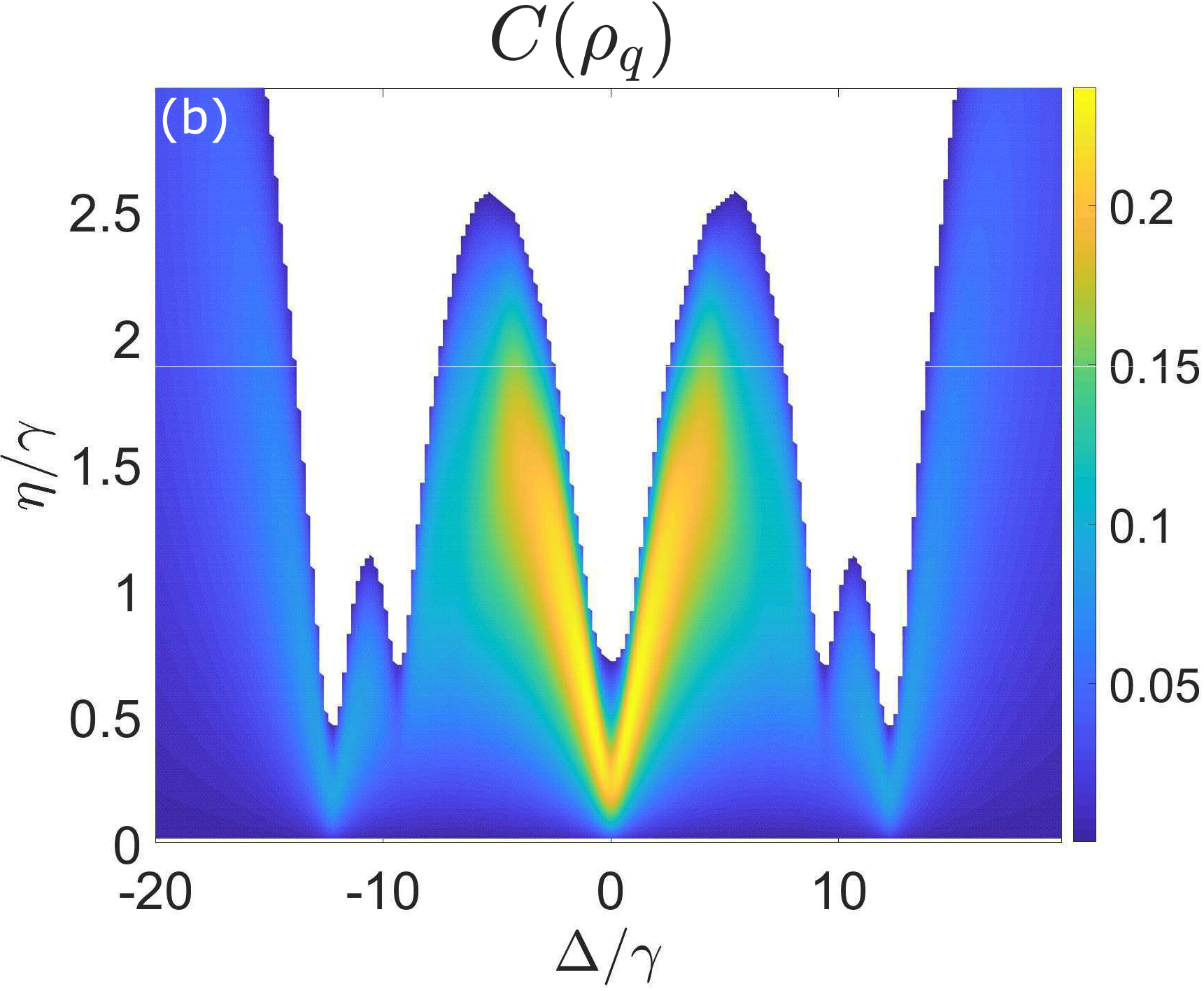}\\
 \includegraphics[width=0.45\linewidth]{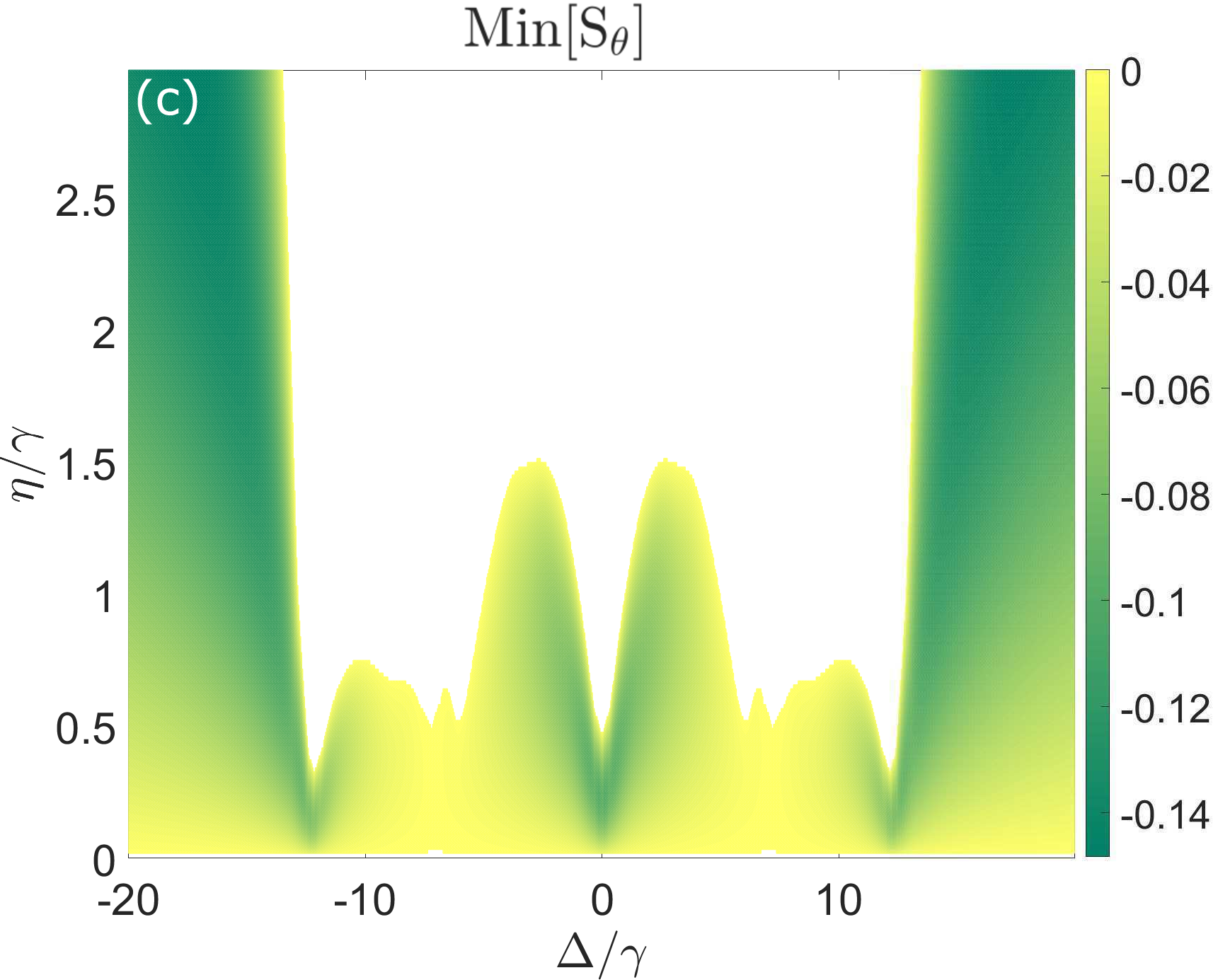}
 \includegraphics[width=0.45\linewidth]{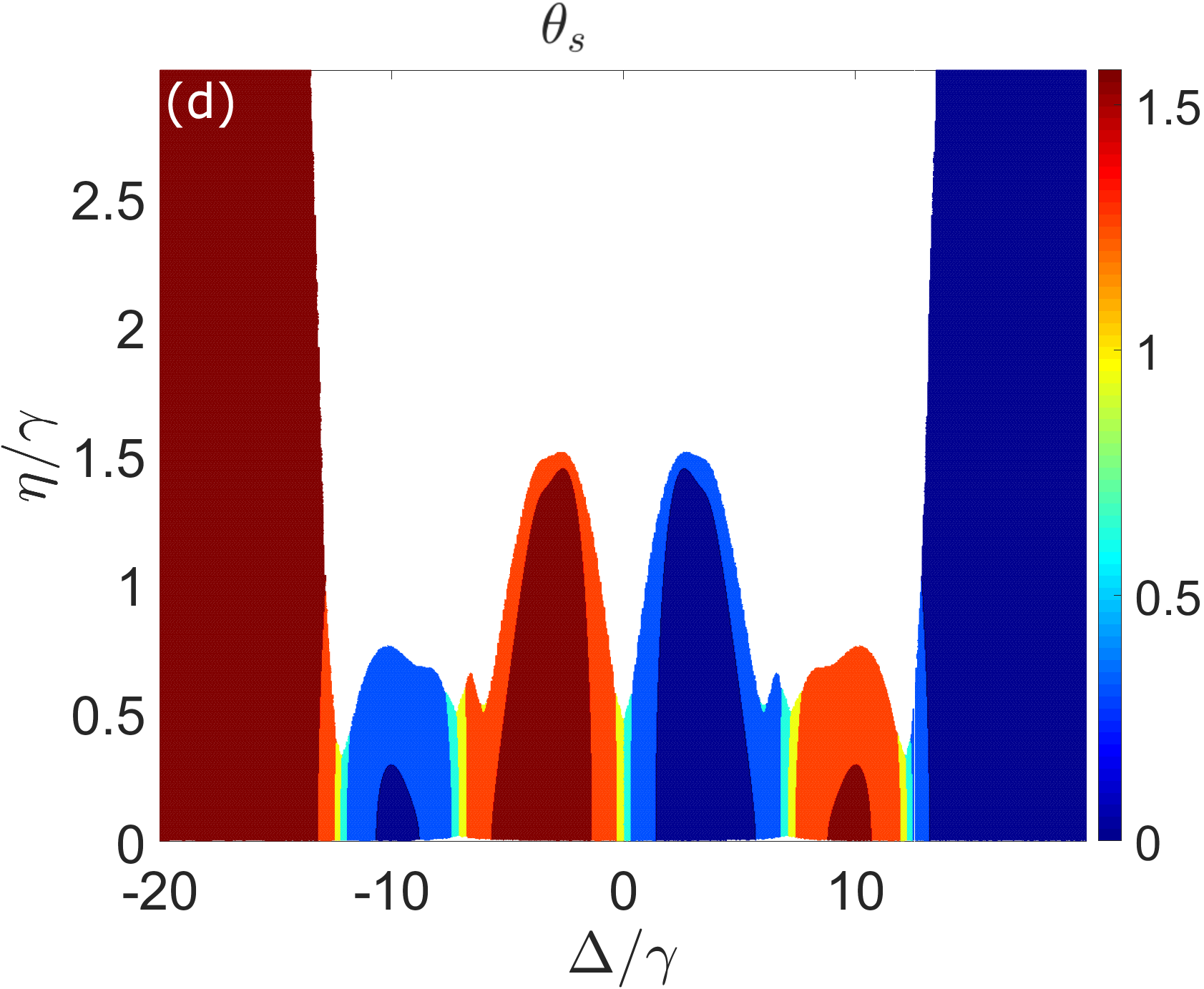}\\

 \caption{(a) The radiance witness $R$, (b) concurrence $C(\rho_q)$ and (c) Min[$S_{\theta}$] are plotted as a function of dimensionless pump strength $\eta/\gamma$ and dimensionless detuning $\Delta/\gamma$. (d) Squeezing angle $\theta_s$ is plotted on the plane of the $\Delta/\gamma$ and $\eta/\gamma$. The rest of the parameters are $g_1=10\gamma,g_2=10\gamma,\kappa=0.5\gamma$ and $\chi=0$.  The blank areas in (b) corresponds to zero concurrence. }
  \label{fig:1}
 \end{figure}

 In  Fig. \ref{fig:1}(a),  the radiance witness $R$ is plotted as a function of $\Delta$ and  $\eta$. Figure \ref{fig:1}(b) shows that the concurrence for the same parameters as in  Fig. \ref{fig:1}(a)  is finite   for small $\eta$  near $\Delta/\gamma=0,\pm \frac{\sqrt{6}g}{2}$. At these parameter regimes. $R$ is quite large as can be seen from  Fig. \ref{fig:1}(a).  These results can be explained by the rate equations as given in equation (\ref{rate eq.}) for transitions between dressed states as we discuss below. 

For weak driving, one can resort to an effective two-state analysis for the transition $|\psi_0^{(0)}\rangle\rightarrow |\psi_{0}^{(2)}\rangle$ which is resonant at $\Delta=0$,  since the other two dressed states $|\psi_{\pm}^{(2)}\rangle$ are energetically away from  $|\psi_{0}^{(2)}\rangle$ by a large gap $\pm\frac{\sqrt{6}g}{2}= \pm 12.2$. Under this two-state approximation(TSA), the equation (\ref{rate eq.}) reduces to
 \begin{equation}
     \dot\rho_{20}=-(i \tilde{\delta}+ \tilde{\kappa})\rho_{20}+i\tilde{\Omega}(\rho_{00}-\rho_{22})
 \end{equation}
 where $\tilde{\kappa} = \frac{\kappa+2}{3}$, $\tilde{\Omega} = \sqrt{2/3} \eta$, $\tilde{\delta} = 2 (\Delta - \omega_{20}/2)$, 
$\rho_{20}=\langle\psi_0^{(2)}|\rho|\psi_0^{(0)}\rangle$, $\rho_{00}=\langle\psi_0^{(0)}|\rho|\psi_0^{(0)}\rangle$, $\rho_{22}=\langle\psi_0^{(2)}|\rho|\psi_0^{(2)}\rangle$. Here $\omega_{20} = \omega_{\rm dr}^{(20)} - \omega_{\rm dr}^{(00)} =0 $ with $\omega_{\rm dr}^{(20)}$ and $\omega_{\rm dr}^{(00)}$ being the dressed frequency of the states $|\psi_0^{(2)}\rangle$ and $|\psi_0^{(0)}\rangle$, respectively; as given in the table \ref{tab:my_label}. The solutions are given by the well-known expressions \cite{cohen1998atom}
\begin{align*}
\rho_{20}&=\frac{(\tilde{\delta}+i{\tilde\kappa})}{\tilde{\Omega}}\frac{s}{s+1}\\
\rho_{22}&=\frac{s}{2(s+1)}
\end{align*}
where the saturation parameter $s=\frac{\tilde{\Omega}^2/2}{\tilde{\delta}^2+\tilde\kappa^2}$. 
The reduced field density matrix under TSA can be written as $\rho_F=\rm{Tr_{\rm atom}[\rho]}=\frac{\rho_{22}|2\rangle\langle2|}{3}+\frac{2\rho_{22}|0\rangle\langle0|}{3}-\frac{\rho_{20}|2\rangle\langle0|}{\sqrt{3}}-\frac{\rho_{02}|0\rangle\langle2|}{\sqrt{3}}$.
So, we have $\langle a^{\dagger} a \rangle_2 = \rm{Tr_{\rm field}[\rho_F]} = 2 \rho_{22}/3$. If we consider only one qubit inside the cavity, then the single-qubit dressed states in the $n$-photon($n \ge 0$)sector are 
\begin{align*}
\mid \psi_{+}^{(n)} \rangle &= 
\cos{\theta}|g,n+1\rangle+ \sin{\theta}|e,n\rangle\\
\mid \psi_{-}^{(n)} \rangle &=
-\sin{\theta}|g,n+1\rangle+ \cos{\theta}|e,n\rangle\;
\end{align*}
where $\tan(2\theta)=-\frac{2g\sqrt{n+1}}{\delta_c}$ with $\delta_c$ being the qubit-cavity detuning. The corresponding dressed frequencies are  $\omega_{\pm}^{(n)}= - \delta_c/2 \pm \sqrt{\delta_c^2 + 4 (n+1) g^2}/2$. Now, two-photon drive can couple the state $\mid g, 0 \rangle$ to $\mid \psi_{\pm}^{(1)}\rangle$ states. Since for our numerical work we have assumed $\delta_c = 0$, 
the dressed frequencies of $\mid \psi_{\pm}^{(1)}\rangle$ are $\mp \sqrt{2} g $. So, when the drive is on resonance with the cavity field, that is, $\Delta = 0$, the single-qubit transitions $\mid g, 
 0 \rangle \rightarrow \mid \psi_{\pm}^{(1)}\rangle $ are off-resonant by $\mp g/\sqrt{2}$ which are $\mp 7.1$ for $g=10$. On the other hand, when the drive is tuned to  $\mp g/\sqrt{2}$, there will be resonant transitions $\mid g, 
 0 \rangle \rightarrow \mid \psi_{\pm}^{(1)}\rangle $ and so the single qubit quantum dynamics will dominate over the correlated two-qubit dynamics. Like two-qubit case, if we consider the dressing of the two states $\mid g, 0\rangle $ and $\mid \psi_{+}^{(1)}\rangle$ or $\mid \psi_{-}^{(1)}\rangle$  by the drive, we can similarly calculate analytically the average photon number $\langle a^{\dagger} a \rangle_1$. The result is $\langle a^{\dagger} a \rangle_1 = s_1/2(s_1+1)$ where $s_1 =\frac{2\eta^2}{4 (\Delta \pm g/\sqrt{2})^2+\kappa^2}$. Writing the radiance $R$ of equation (\ref{R_TSA}) in the form $R = Q - 1$ where $Q = \langle a^{\dagger} a \rangle_2/(2 \langle a^{\dagger} a \rangle_1) $, we can express the quantity $Q$  as 
 \begin{equation}
     Q=\frac{1}{3}\left[\frac{\eta^2+2(\Delta \pm \frac{g}{\sqrt{2}})^2+\frac{\kappa^2}{2}}{\eta^2+12\Delta^2+\frac{(\kappa+2)^2}{3}}
     \right]
     \label{R_TSA}
 \end{equation}
If $Q > 1$ we have hyperradiance. For $\eta <\!< \kappa$, $\Delta \simeq 0$ and $g > 2$, the above expression shows that $Q > 1$. On the other hand, for $|\Delta| \simeq g/\sqrt{2} $, $Q < 1$ we have sub-radiance. The quantum dynamics at $\Delta \simeq 0$  is dominated by two-qubit transition $\mid \psi_0^{(0)} \rangle \rightarrow \mid \psi_{0}^{(2)}\rangle$ and $\Delta \simeq \pm g/\sqrt{2}$ regime is dominated by single-qubit transitions $\mid g, 0 \rangle \rightarrow \mid \psi_{\pm}^{(1)} \rangle $ due to the two-photon drive.  So, we can infer that whether hyperradiance or sub-radiance will appear depends on the interplay between the two-qubit and single-qubit dressed state transitions. For $g/\sqrt{2} >\!> |\Delta| \ge 0 $, hyperradiance will be prominent while for $\sqrt{6}g/2 >\!> |\Delta| \simeq g/\sqrt{2} $ sub-radiance will appear due to the dominant single-qubit transitions.

Now, to explain the qubit-qubit entanglement in the small $\eta$ regime, we can consider that the drive further dresses the two dressed states under TSA. The such doubly dressed (dd) states have the eigen energies $-\tilde{\delta}/2 \pm \sqrt{\tilde{\delta}^2 + \tilde{\Omega}^2}/2$.  
For the transition $\mid \psi_0^{(0)}\rangle \rightarrow \mid \psi_{0}^{(2)}\rangle$, 
$\tilde{\delta} = \Delta$. When $\Delta = 0$, the two  dd states: (i) $\mid\pm \rangle_{{\rm dd}} = [|\psi_{0}^{(2)}\rangle \pm |\psi_0^{(0)}\rangle]/\sqrt{2}$. On taking projection of these two states onto the zero photon state, one obtains two two-qubit entangled states of the form $c_{gg} \mid g g\rangle \pm c_{ee} \mid e e\rangle$. The probability of such entangled states will be given by $|\rho_{20}|$.  So, the concurrence is expected to be proportional to $|\rho_{20}|=\frac{\tilde{\Omega}(\sqrt{\tilde{\delta}^2+\tilde{\kappa}^2)}}{\tilde{\Omega}^2/2+ \tilde{\delta}^2+\tilde{\kappa}^2}$  which clearly reveals the dispersive nature of the concurrence which will vanish for the limit $\tilde{\Omega} \rightarrow \infty$ which amounts to $\eta \rightarrow \infty $ and also for the limit $\Delta \rightarrow \pm \infty$.

 Next, to discuss nonclassicality of the field, we plot the minimum of the squeezing-parameter Min$[S_\theta]$ as a function of $\Delta/\gamma$ and $\eta/\gamma$ in Fig. \ref{fig:1}(c). There are three significant peaks at $\Delta=0,\pm \sqrt{6} g/2\gamma$ for weak pumping. The corresponding squeezing angle $\theta_s$ is  plotted in Fig. \ref{fig:1} (d). The squeezing angle  depends predominantly on the detuning rather than the pump field strength, as illustrated in Fig. \ref{fig:1} (d). So, one can produce squeezed field with any desired squeezing angle, such as amplitude squeezing or phase squeezing,  by simply adjusting the detuning.  The rest of the parameters in Fig. \ref{fig:1} are $g_1=10\gamma,g_2=10\gamma,\kappa=0.5\gamma$ chosen to fulfill the strong coupling CQED limit.
 
\begin{figure}
    \centering
   \includegraphics[width=0.30\linewidth]{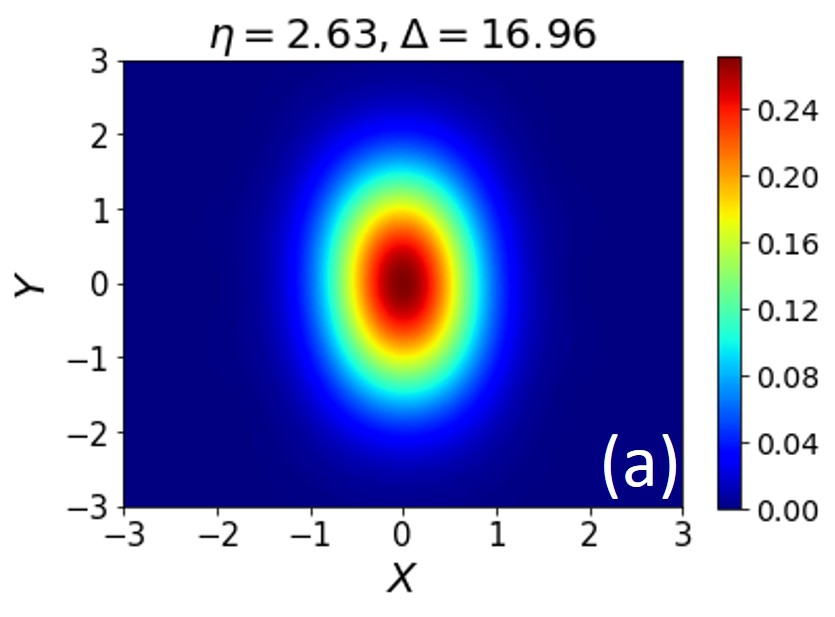} 
    \includegraphics[width=0.30\linewidth]{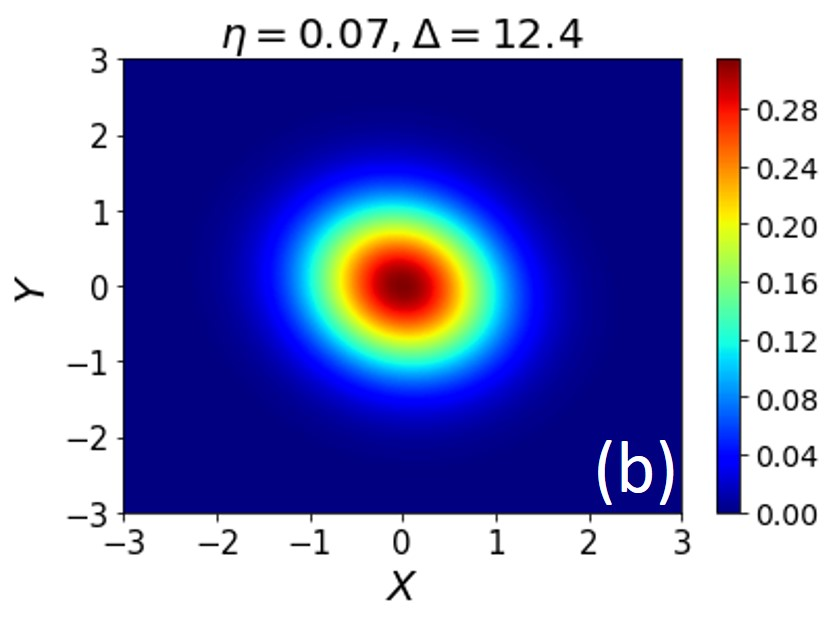} 
    \includegraphics[width=0.30\linewidth]{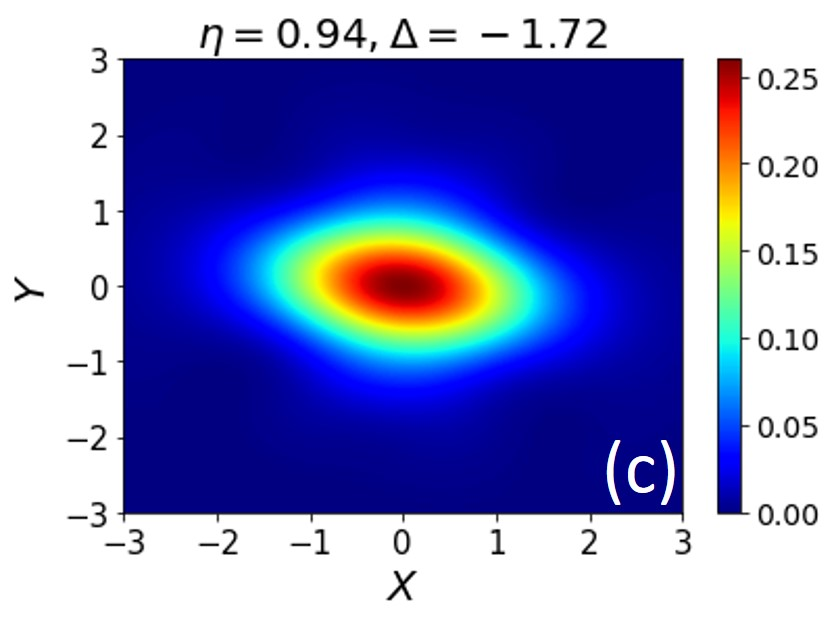}\\
    \includegraphics[width=0.30\linewidth]{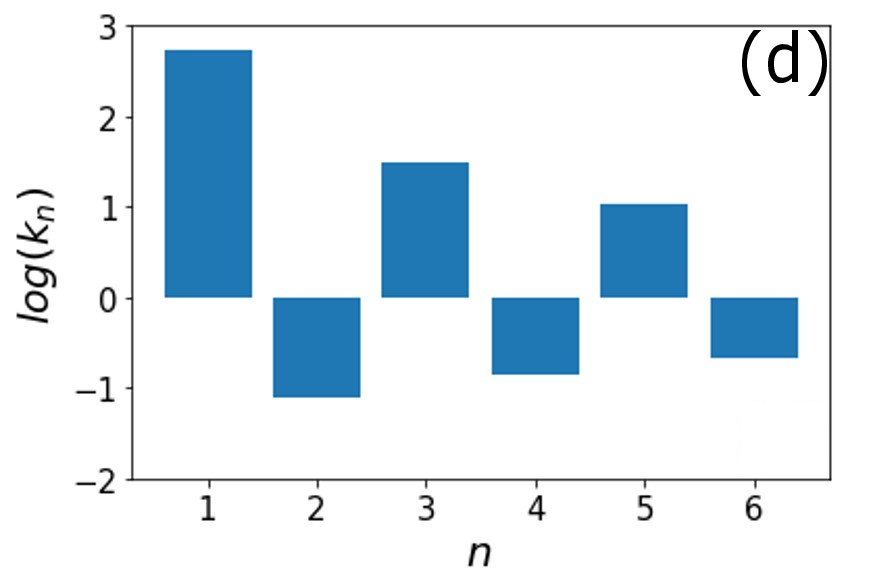} 
    \includegraphics[width=0.29\linewidth]{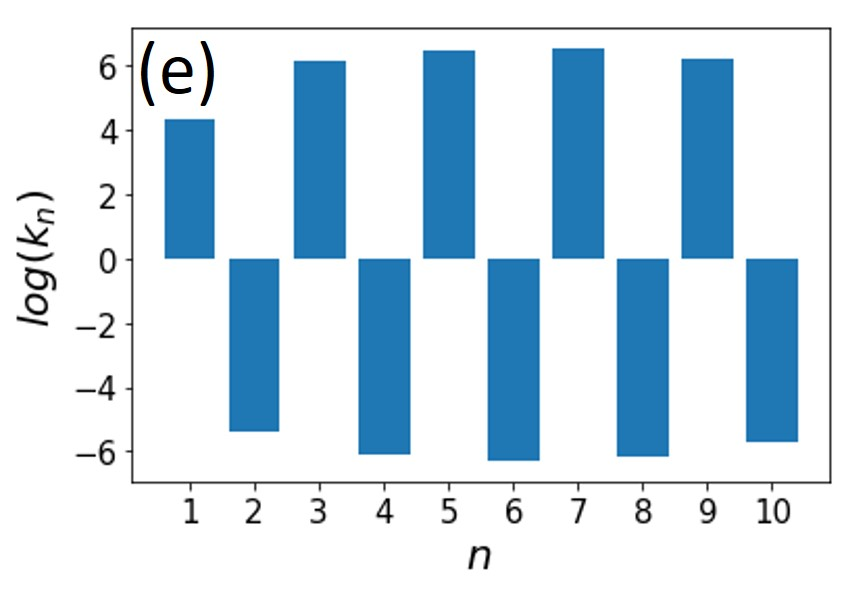} 
    \includegraphics[width=0.31\linewidth]{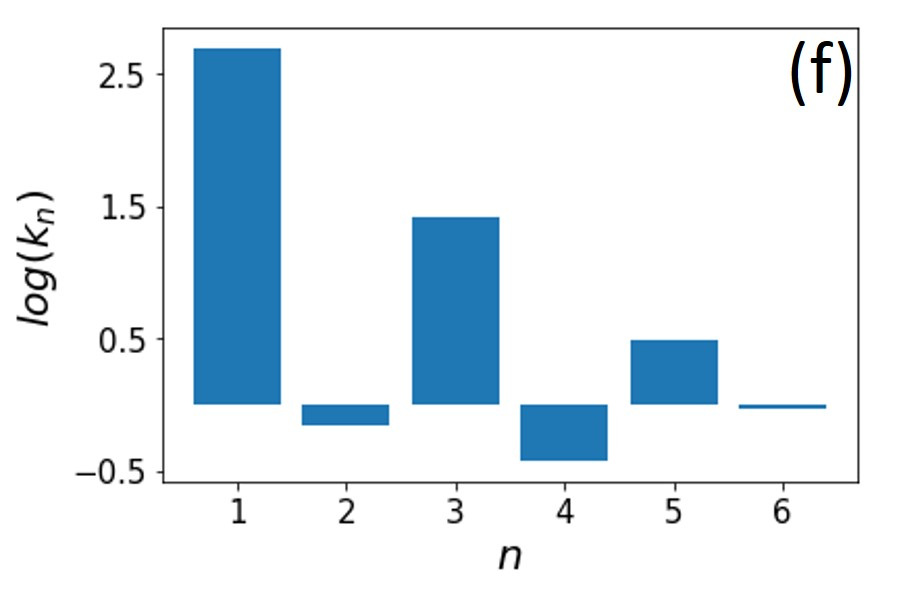}
    
    \caption{The density plots of Wigner functions ((a), (b) and (c)) and the logarithm of corresponding Klyshko's parameter ($\rm{log}K_n$) ((d), (e) and (f)) are depicted for three sets of $\eta/\gamma$ and $\Delta/\gamma$ values as mentioned at the top of the each sub-figures of (a), (b) and (c) chosen from Fig. \ref{fig:1}(c). The other parameters are same as   in Fig. \ref{fig:1}}
    \label{fig:2}
\end{figure}

We next display the Wigner functions of the cavity field in Figs. \ref{fig:2}(a-c) for three typical parameter sets chosen from the Fig. \ref{fig:1}(c) to have strong squeezing. In contrast to the wider range of fluctuations as Fig. \ref{fig:2}(b) shows, the Wigner functions of Figs. \ref{fig:2}(a) and Fig. \ref{fig:2}(c) show elliptical profiles in the parameter regime of squeezing, resulting in the elongation of the distribution in one particular direction. Figures. \ref{fig:2}(a) and \ref{fig:2}(c) exhibit squeezing along almost vertical and horizontal directions, respectively, depending on the various pump field detunings. 

To illustrate the nature of photon distributions, we plot the  Klyshko's criteria distribution $K_n$ in Figs. \ref{fig:2}(d–f). The odd-even oscillations of $K_n$ alternates up and down mirroring the expected even-odd photon distribution of squeezed states. It is important to note that unlike \ref{fig:2}(b) that corresponds to the resonant case i.e $\Delta=\pm \sqrt{6}g/2\gamma$, the off-resonant cases in Fig. \ref{fig:2}(a) and Fig.\ref{fig:2}(c) show a rapid decay of the even-odd oscillations with the increase of photon number. It means that the squeezed states generated in the off-resonant cases involve mainly two-photon excitations where in resonant cases higher order even number transition processes may take place.

\begin{figure}
    \centering
   \includegraphics[width=0.45\linewidth]{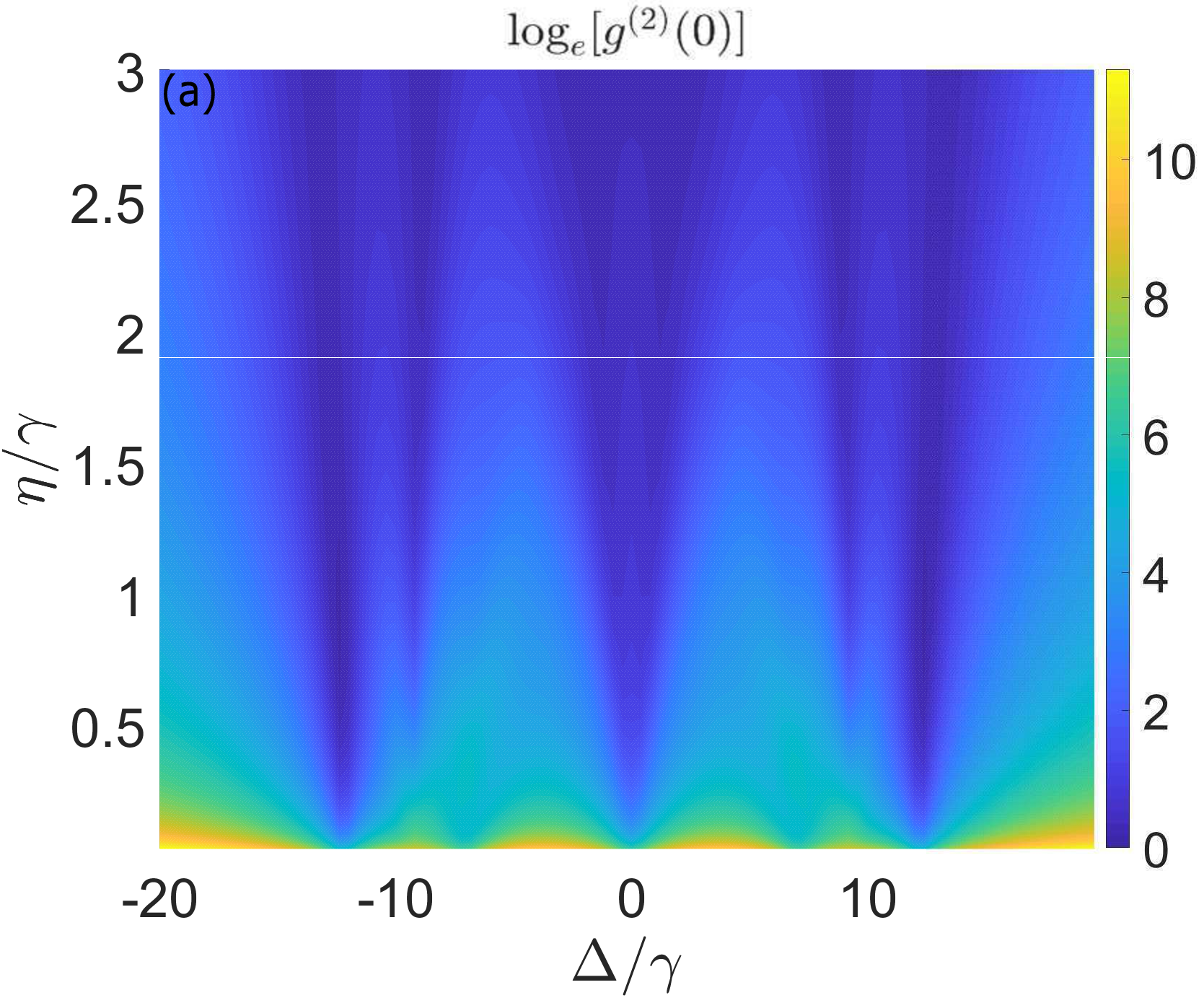} 
    \includegraphics[width=0.45\linewidth]{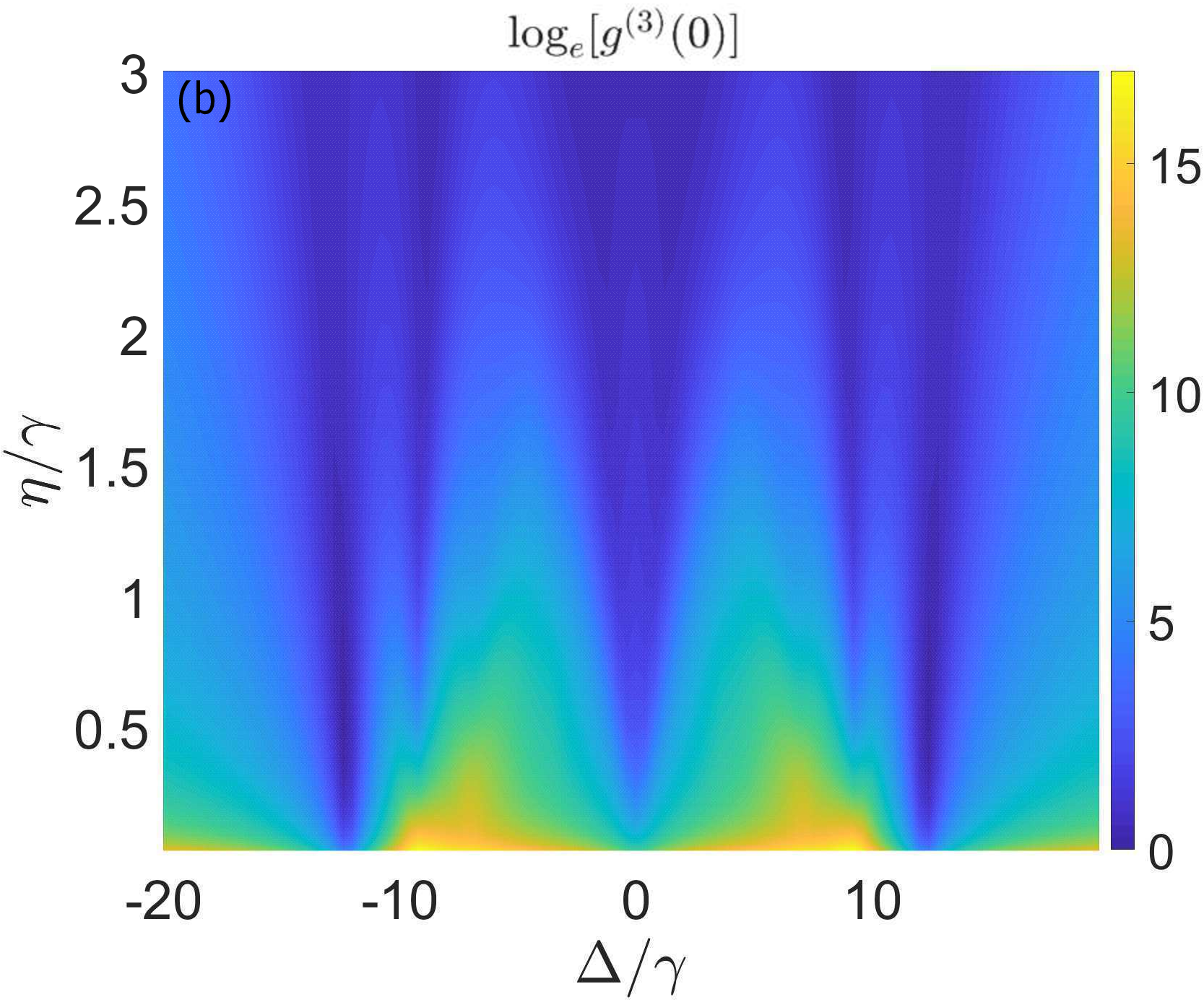} 
    
    \caption{The second and third order HBT correlation functions  $\rm{log}_{e}g^{(2)}(0)$ (a) and $\rm{log}_{e}g^{(3)}(0)$ (b) are plotted as a function of $\Delta/\gamma$ and $\eta/\gamma$. The  other parameters are same as in Fig. \ref{fig:1}. As neither of $g^{(2)}(0)$ and $g^{(3)}(0)$ is less than 1, one- and two- photon blockades do not exist. }
    \label{fig:3}
\end{figure}

In order to understand the $n$-photon blockade, $\log_{e}g^{(n)}(0)$ ($n=2,3$) is plotted in the plane of $\Delta/\gamma$ and $\eta/\gamma$ in Figs. \ref{fig:3}(a) - (b) which show that none of the correlations goes below unity ($\log_{e}g^{(n)}(0)<0$). So it is impossible to find a single-photon or two-photon blockade in this entire parameter range. The reason for the absence of single photon blockade can be explained from the dressed-state picture. The transition from $|\psi_0^{(0)}\rangle \rightarrow |\psi_{\pm}^{(1)}\rangle$ is not allowed as they are not coupled by two-photon transitions. The value of two and three photon correlations are very large near $\eta=0$. However, in these positions, the average photon numbers are very low making the statistical measurements a difficult task. 

\subsection{The effects of a Kerr medium}

In the foregoing analysis, we have shown that our model does not allow photon blockade which is an intrinsically nonlinear process, although we have used second order nonlinear medium to produce two-photon drive. 

\begin{figure}
    \centering
    \includegraphics[width=0.45\linewidth]{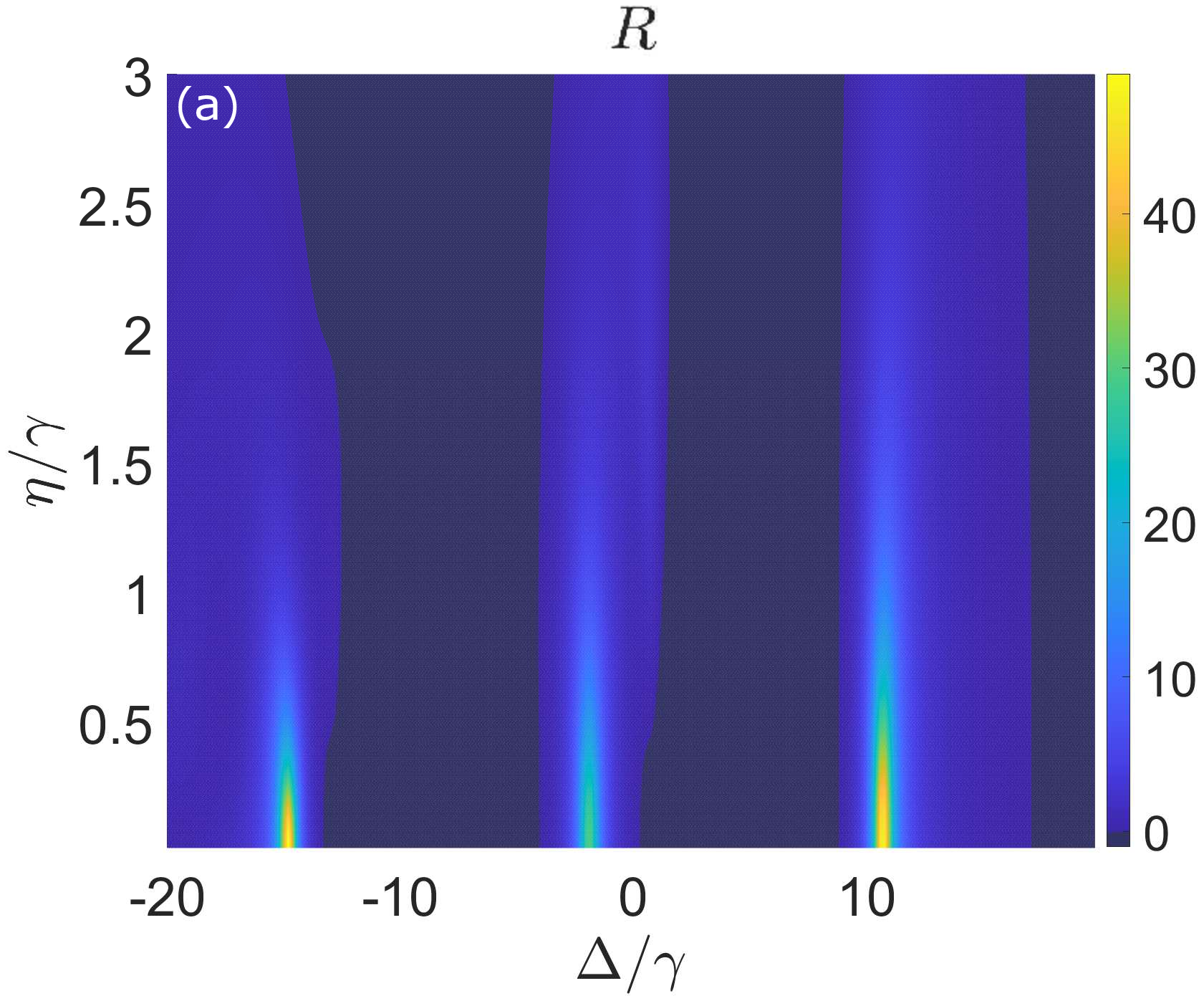}
    \includegraphics[width=0.47\linewidth]{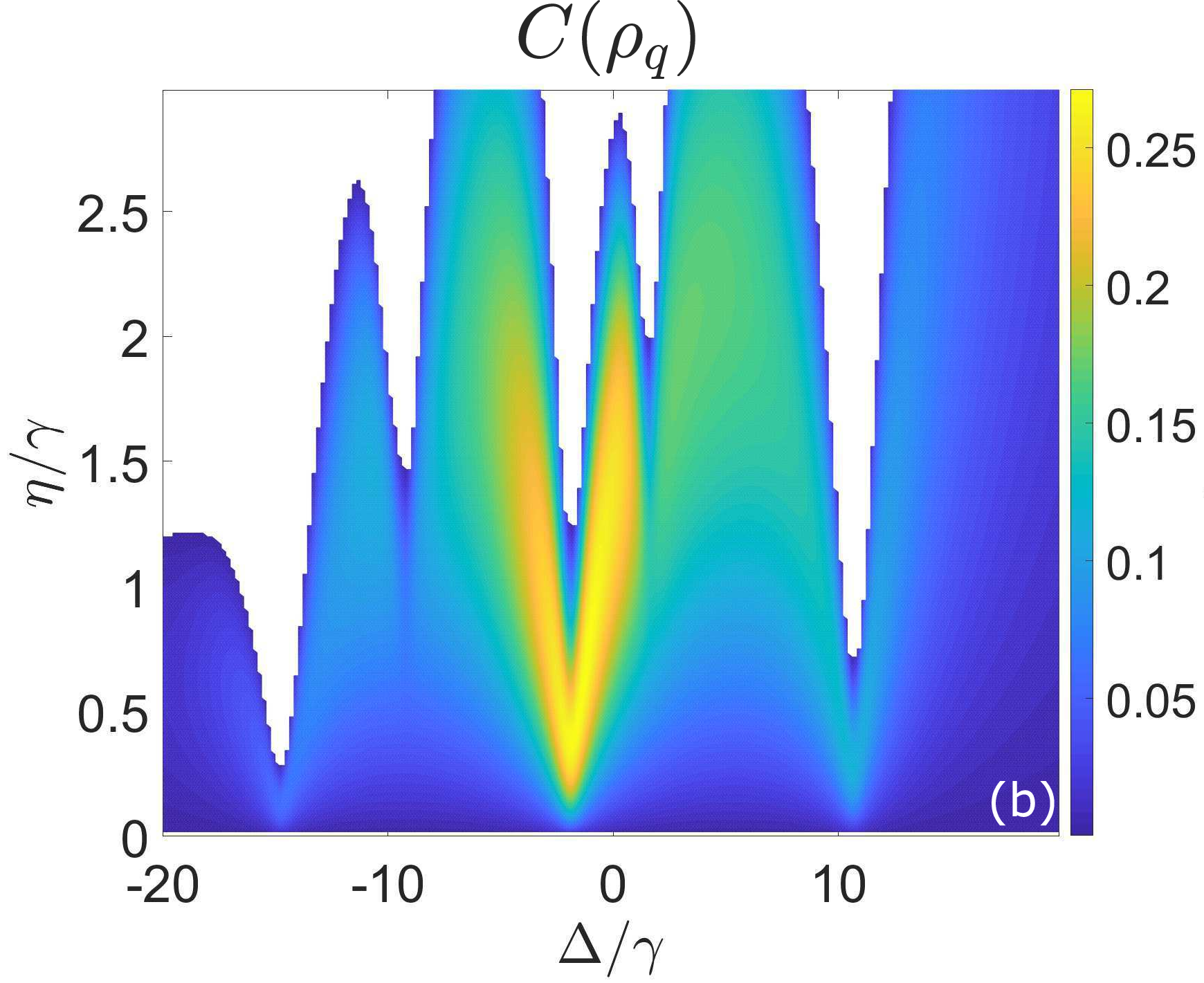}\\
   \includegraphics[width=0.45\linewidth]{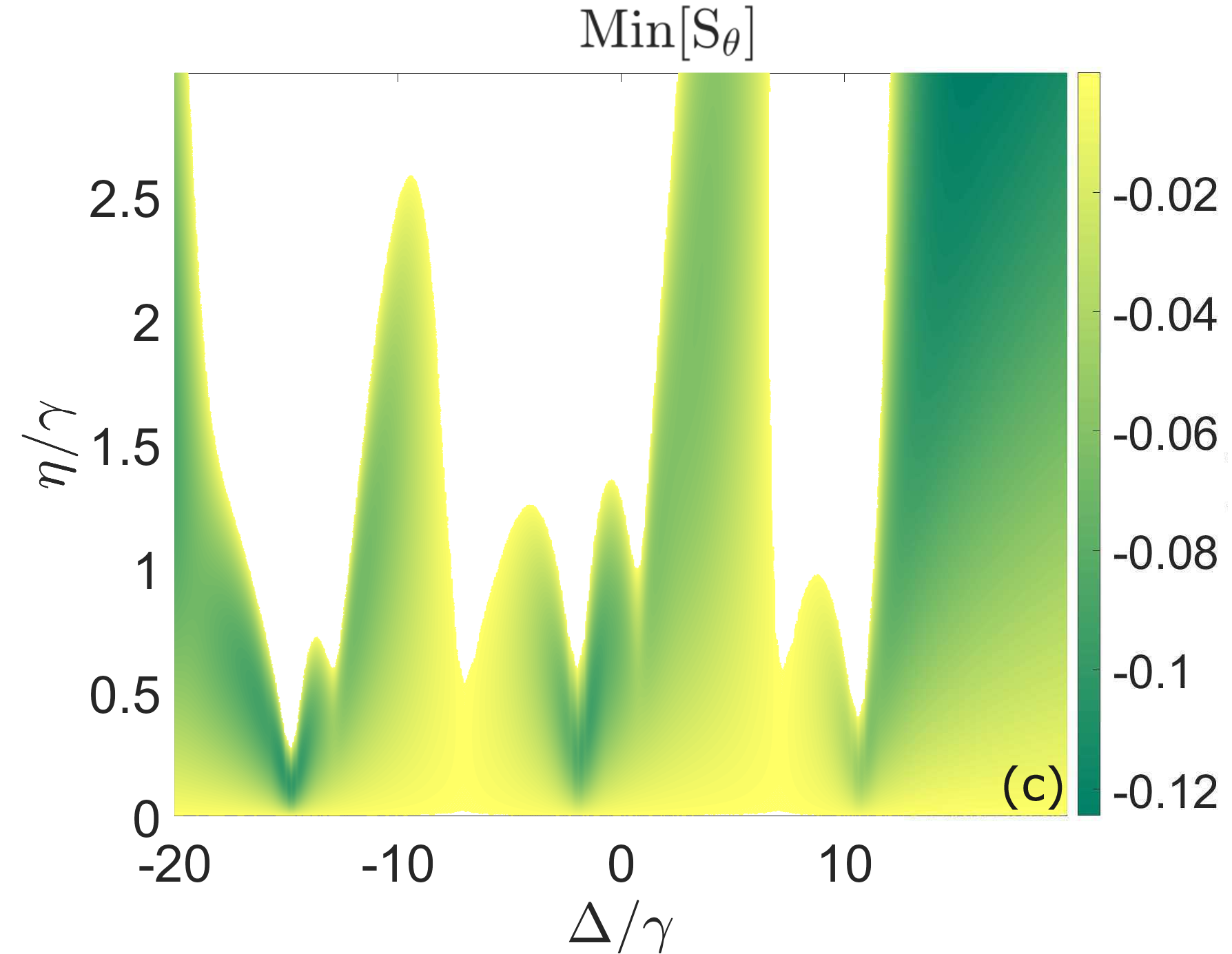} 
    \includegraphics[width=0.45\linewidth]{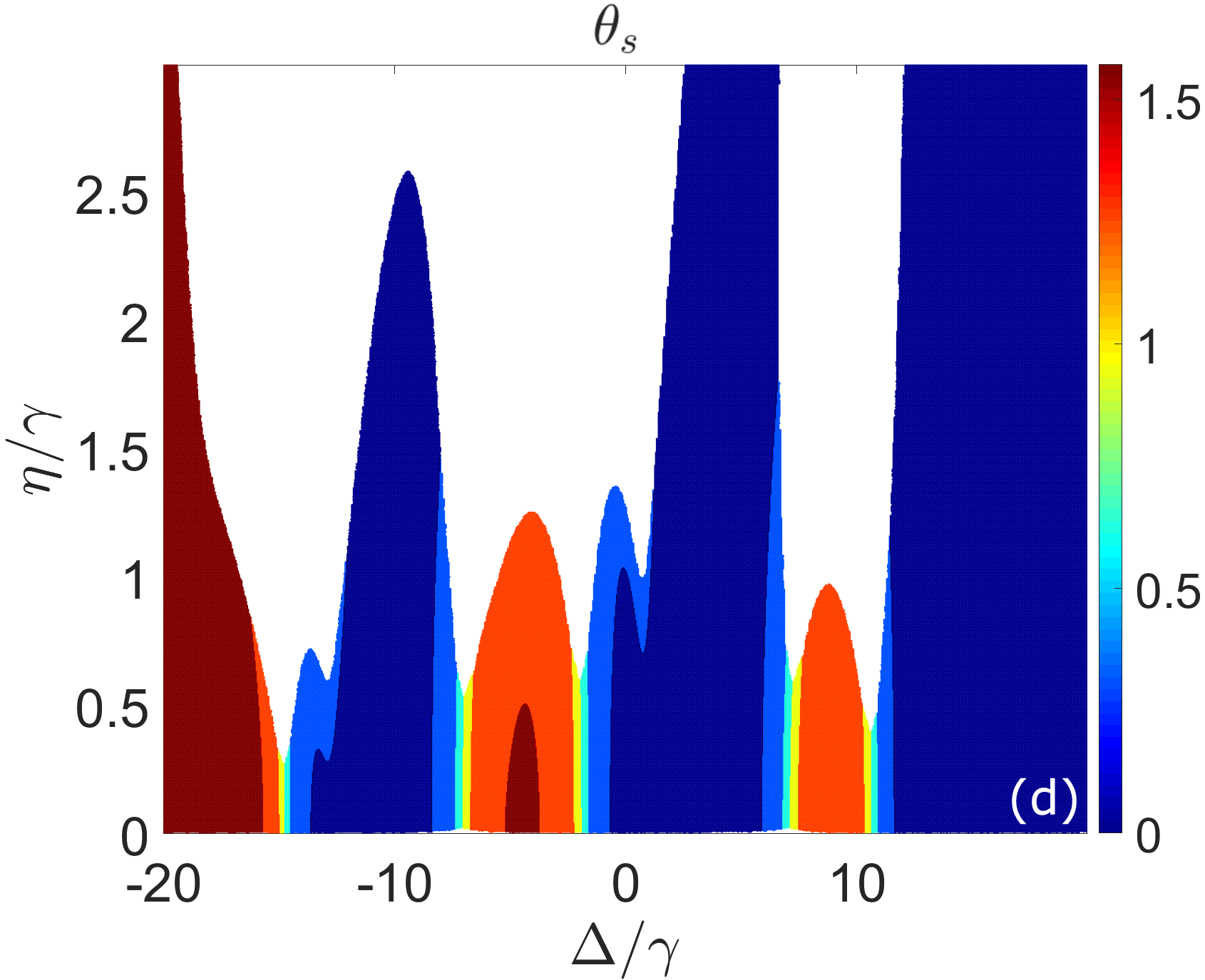}
    
    \caption{(a) $R$, (b) $C(\rho_q)$ and (c) Min[$S_{\theta}$] are plotted as a function of $\eta/\gamma$ and $\Delta/\gamma$. The greener area corresponds to relatively stronger squeezing. (d) Squeezing angle $\theta_s$ is plotted on the plane of the $\Delta/\gamma$ and $\eta/\gamma$. The white area of figure (b) shows $C(\rho_q)=0$. The blank areas in both the figure (c) and (d) indicates that the minimum squeezing parameter Min$[S_{\theta}]>0$. The other parameters are $g_1=10\gamma,g_2=10\gamma,\kappa=0.5\gamma$, $\chi=6\gamma$.}
    \label{fig:4}
\end{figure}

The dressed states and dressed energies are modified due to Kerr nonlinearity. The ground state remains the same as before but the dressed energies change. It is difficult to obtain analytical expressions for dressed energies and dressed states for finite strength of Kerr nonlinearity $\chi$. For $\chi = 6$ and $g = 10 $, the dressed energies are   
$E^{20'}_{dr}=\hbar \omega^{20'}_{dr}=3.80  $, $E^{2+'}_{dr}=\hbar \omega^{2+'}_{dr}=29.55$ and $E^{2-'}_{dr}=\hbar \omega^{2-'}_{dr}=-21.35$ which  correspond to $\hbar\omega^{20} = 0$, $\hbar\omega^{2\pm} = \pm \sqrt{6} g $, respectively, when $\chi = 0$. It is known  that the Kerr-nonlinearity  causes photon blockade \cite{PhysRevA.104.053718,ashefas2022kerr}. In Fig. \ref{fig:4}(a) $R$ is plotted as a function of $\eta/\gamma$ and $\Delta/\gamma$. The three peaks obtained correspond to the three transitions in the dressed-state levels. In Fig. \ref{fig:4}(b) the significant cocurrence is obtained at $\frac{E^{2-'}_{dr}}{2}$, $\frac{E^{20'}_{dr}}{2}$ and $\frac{E^{2+'}_{dr}}{2}$ , the same positions where hyperradiance is obtained. However, the introduction of nonlinearity increases the two-qubit concurrence mainly around the resonant transition at $\Delta/\gamma=\frac{E^{20'}_{dr}}{2}$. 
In Fig. \ref{fig:4}(c) Min$[S_{\theta}]$ and (d) $\theta_s$ are plotted in the plane of $\eta/\gamma$ and $\Delta/\gamma$ with $\chi=6$. Unlike Fig. \ref{fig:1}, this figure shows three peaks at different detunings because the presence of nonlinear medium has modified the dressed state structure. The peaks appear in Fig. \ref{fig:4}(a), (b), (c) and (d) are at $\Delta/\gamma=\frac{E^{2-'}_{dr}}{2},\frac{E^{20'}_{dr}}{2},\frac{E^{2+'}_{dr}}{2}$, respectively. It is because the transitions $|\psi_0^{(0)}\rangle\rightarrow\ |\psi_0^{(2)'}\rangle,|\psi_+^{(2)'}\rangle,|\psi_-^{(2)'}\rangle$ are allowed as can be confirmed by calculating the transition dipole matrix elements between the dressed states. It is to be noted that the Min[$S_{\theta}$] value has almost the same minimum as it is obtained for $\chi=0$.  The minimum squeezing angle $\theta_s$ is usually dependent on the detunings and not on the driving strength as before.

\begin{figure}
 \centering
   \includegraphics[width=0.33\linewidth]{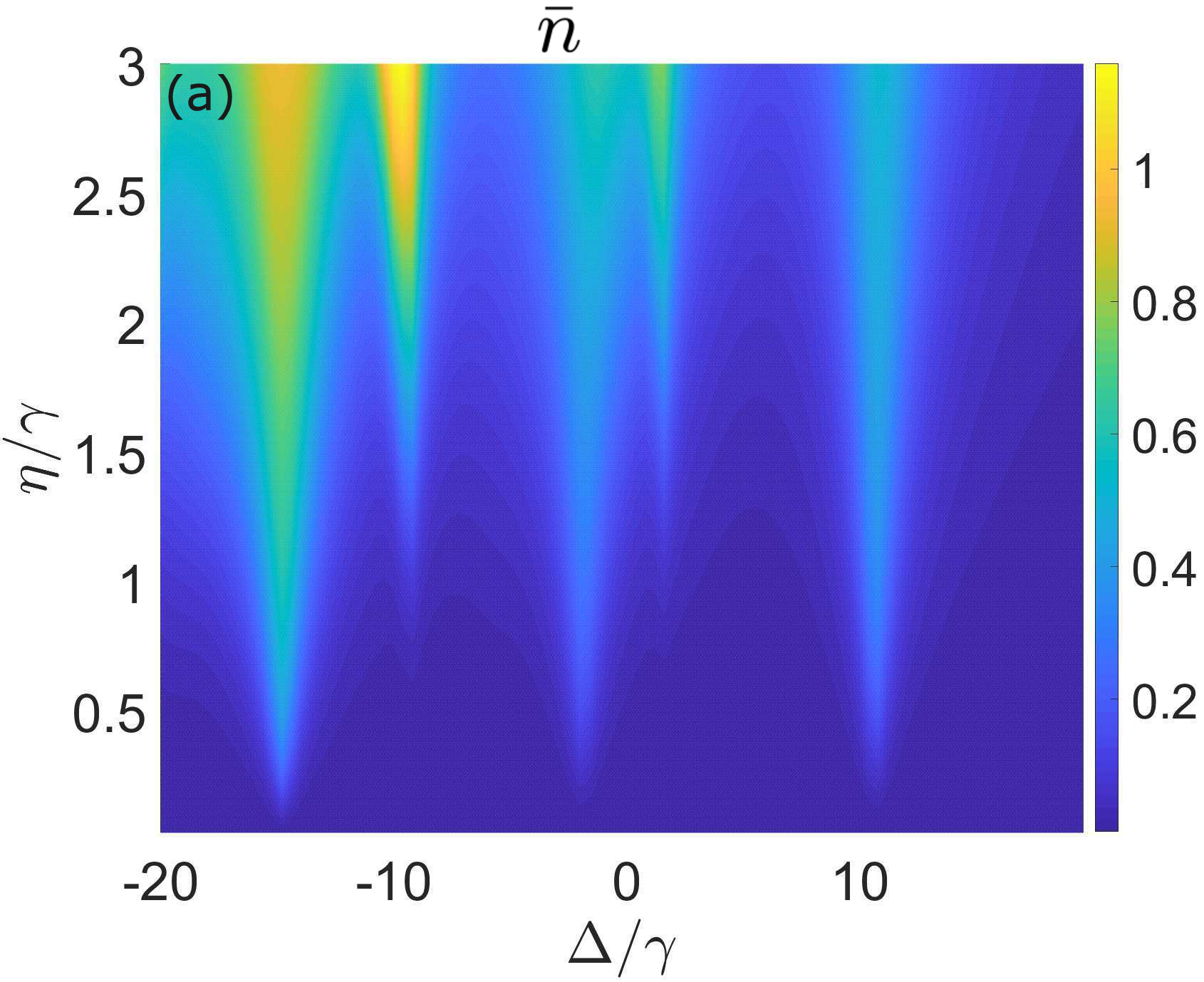} 
   \includegraphics[width=0.3\linewidth]{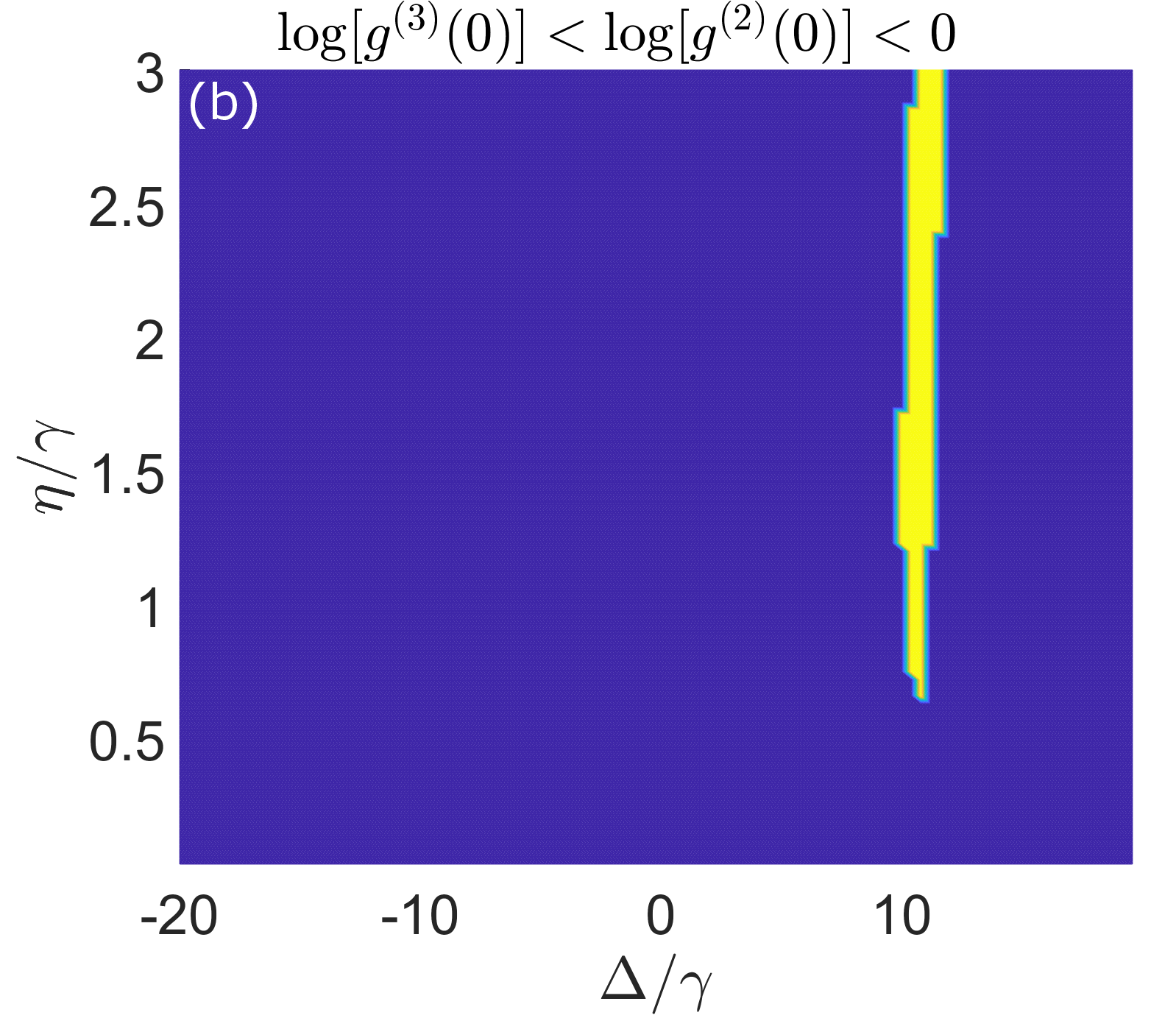}
   \includegraphics[width=0.3\linewidth]{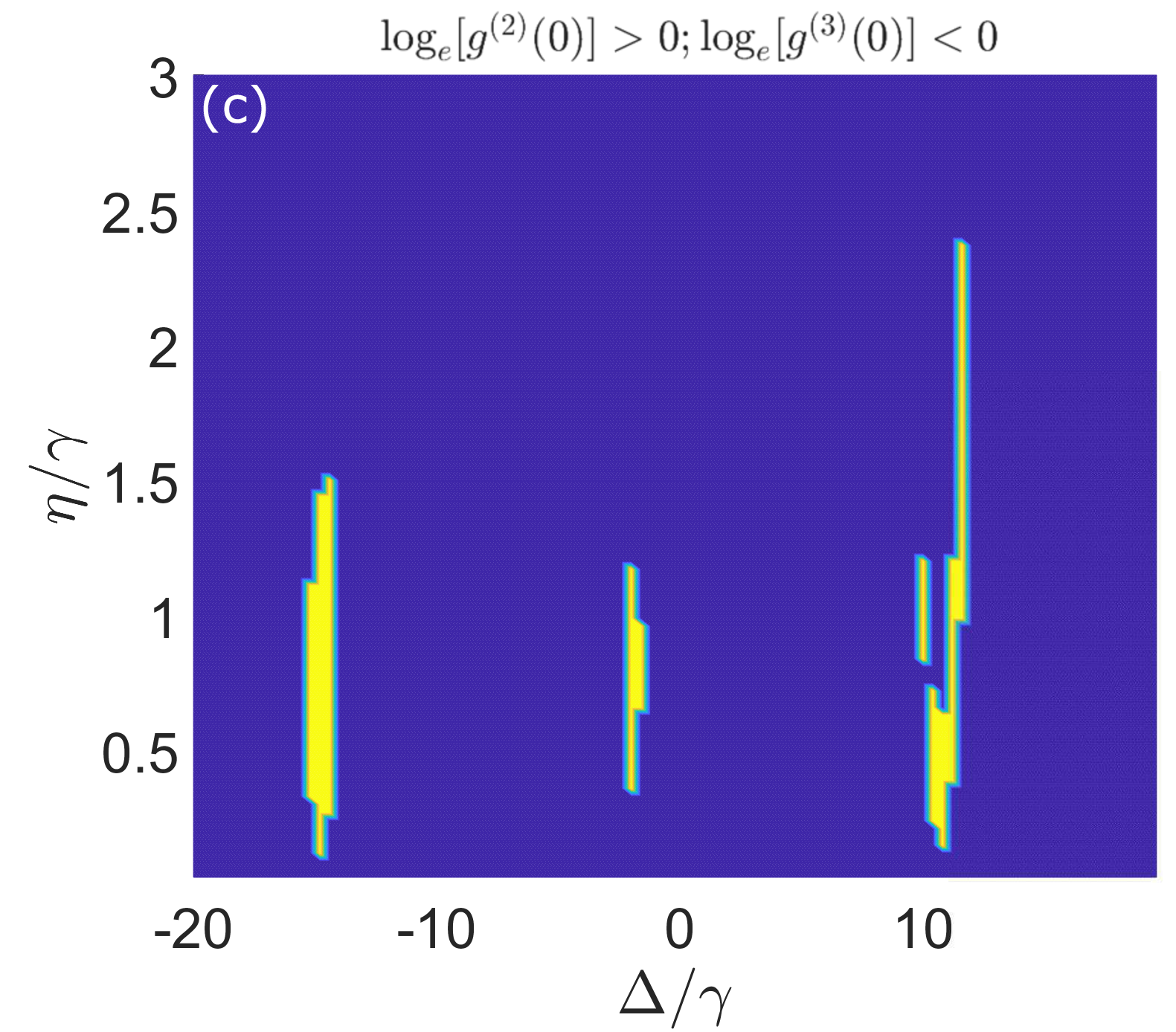}
    \caption{(a) $\bar n$ , (b) the single-photon and (c) two-photon blockade regime (yellow portion) are shown on the plane $\Delta/\gamma$ and $\eta/\gamma$ for the same coupling  with $\chi=6\gamma$. The other parameters are same as in Fig. \ref{fig:4}.}
    \label{fig:5}
\end{figure}

We plot a phase space curve in Fig. \ref{fig:5}(b) and Fig. \ref{fig:5}(c). In Fig. \ref{fig:5}(b), the yellow region describes the condition when both $g^{(2)}(0)$ and $g^{(3)}(0)$ are less than unity. In this region, we also observed that $g^{(3)}(0)<g^{(2)}(0)$. In Fig. \ref{fig:5}(c) the three yellow regions depict the possibility of two-photon blockade characterised by $g^{(3)}(0)<1,g^{(2)}(0)>1$. In the blue region of (b) and (c), the condition concerning the yellow region is not satisfied. Also, we have ensured that in the yellow region, $\bar n$ is appreciable as shown in Fig. \ref{fig:5}(a), otherwise the detection of correlation will be a difficult task.

\begin{figure}
 \centering
   \includegraphics[width=3in]{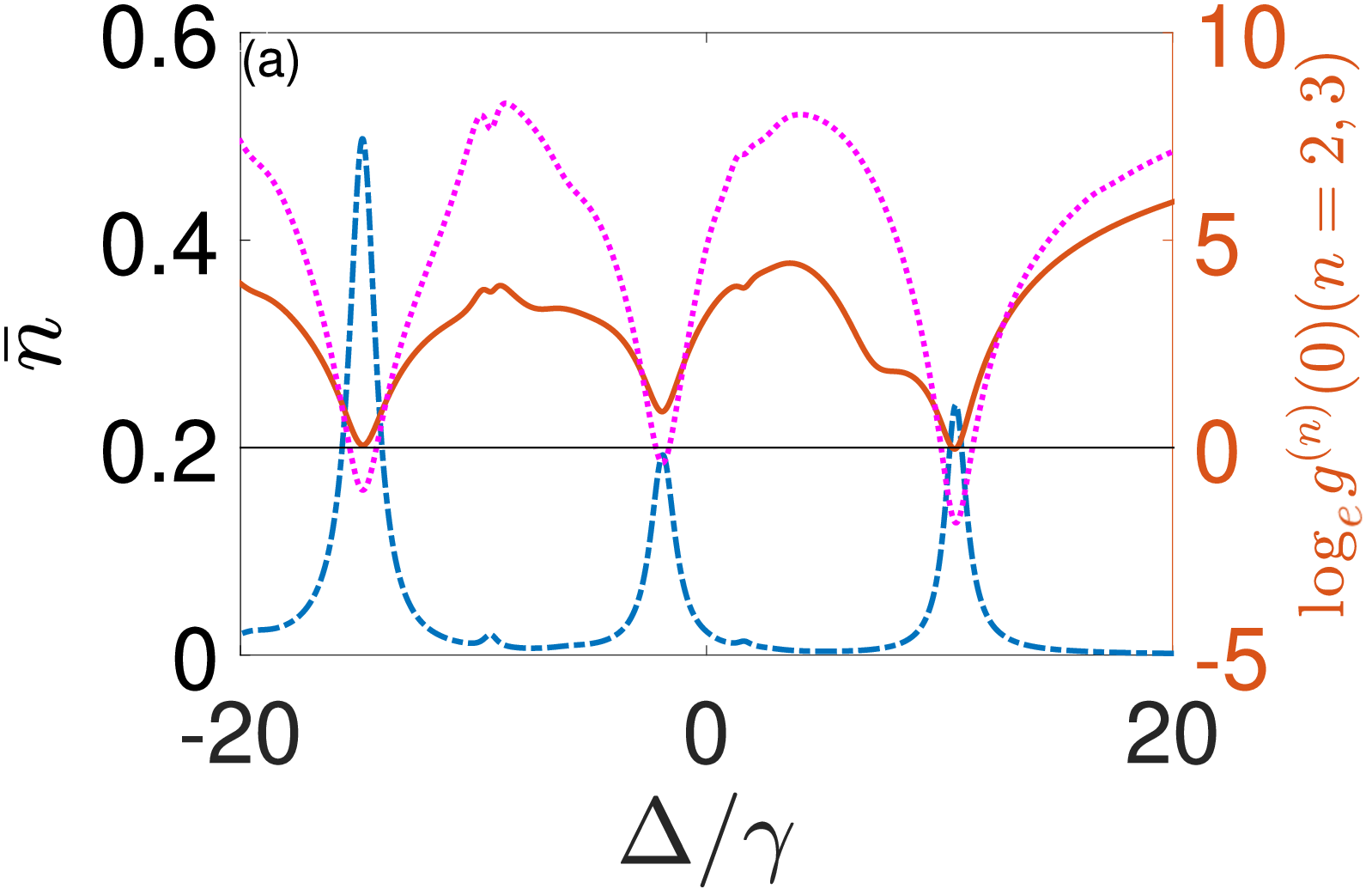} \hspace{0.6em}%
   \includegraphics[width=3in]{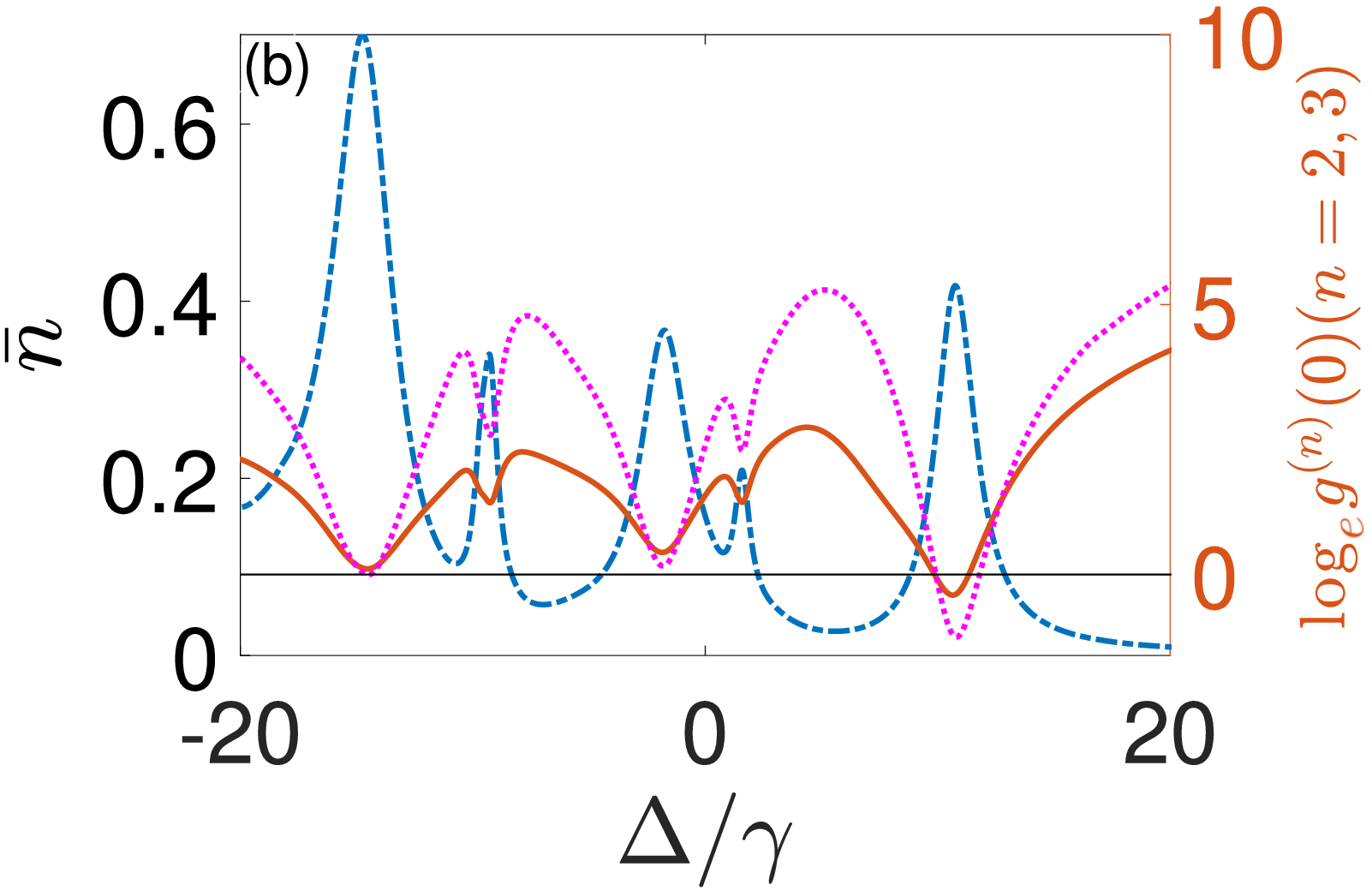}\\
   \hspace{-0.3in}
   \includegraphics[width=2.5in]{ 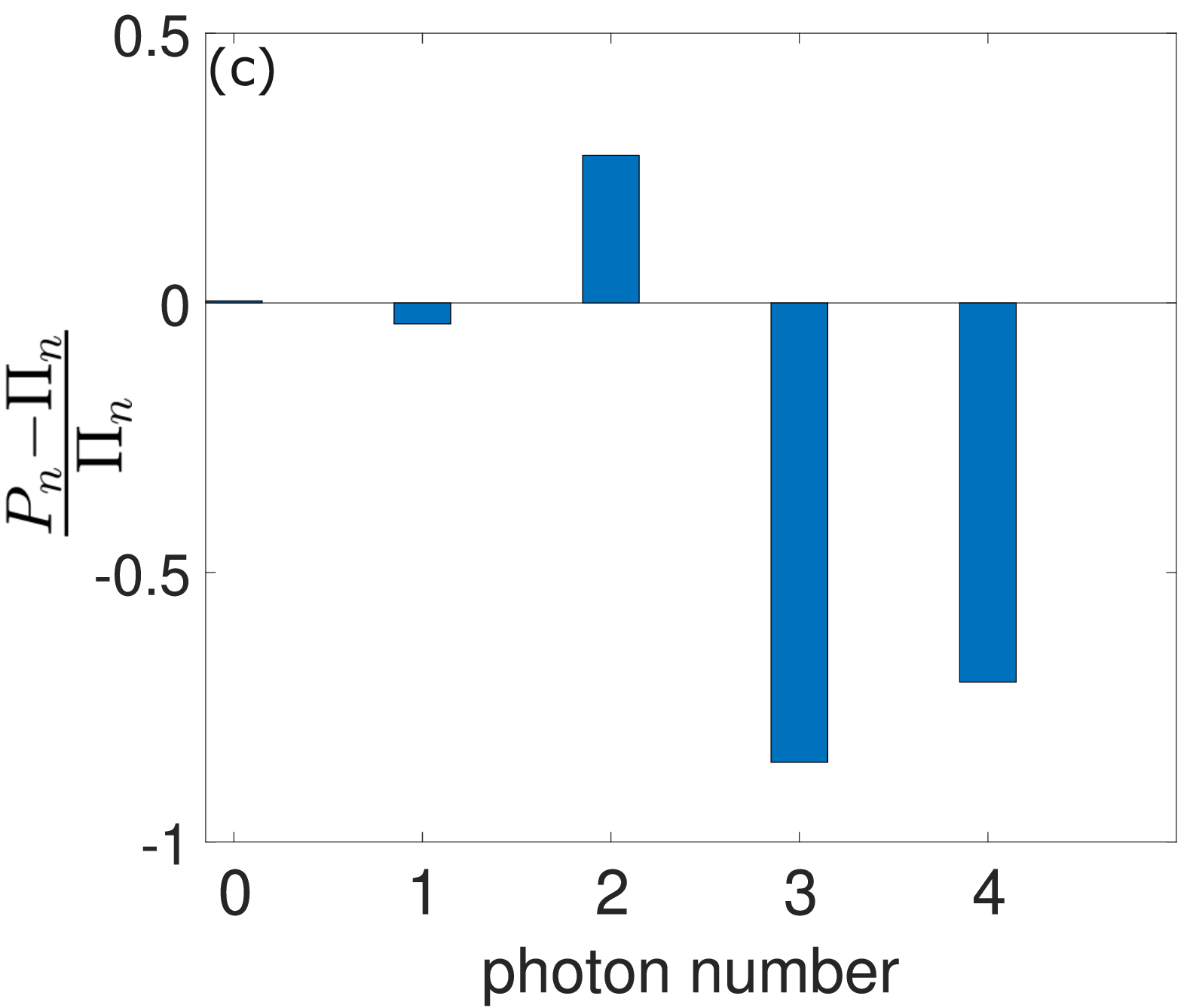} \hspace{5.0em}%
   \includegraphics[width=2.5in]{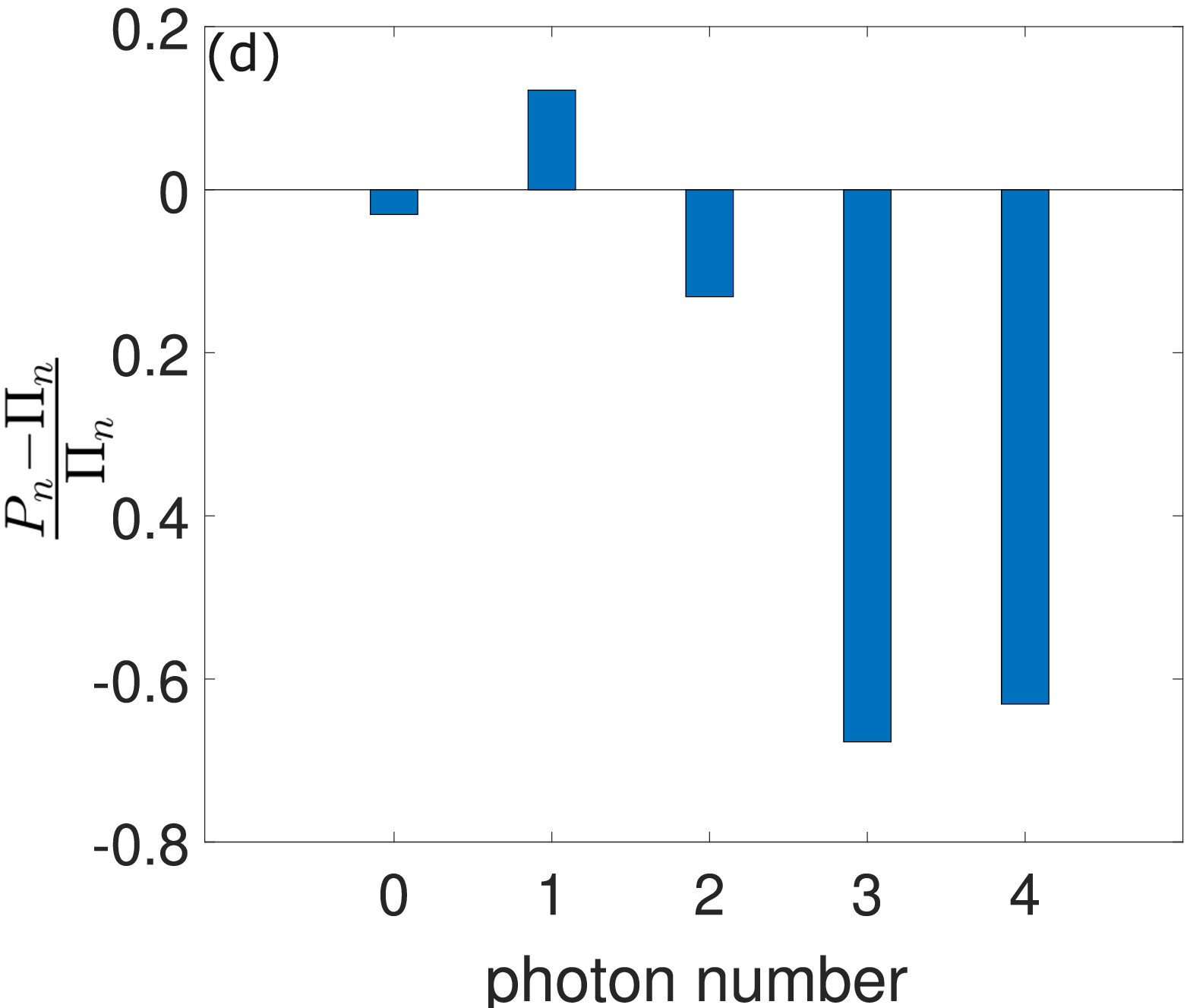}
    \caption{(a) and (b) $\bar{n}$ (dashed-dotted), $\rm{log}_{e}g^{(2)}(0)$ (solid) and $\rm{log}_{e}g^{(3)}(0)$ (dotted) are plotted as a function of $\Delta/\gamma$ for $\eta=0.63\gamma$ (a) and $\eta=1.53\gamma$ (b). The  other parameters are same as  in Fig. \ref{fig:4}. (c) and (d) depict deviations of the photon-number distribution from Poisson distribution when $\Delta/\gamma=\frac{E^{2-'}_{dr}}{2}$. (a) and (c) show two-photon blockade and (b) and (d) show a single-photon blockade at detuning $\frac{E^{2-'}_{dr}}{2}$.}
    \label{fig:6}
\end{figure}

To ensure the existence of $n$-photon blockade, we choose the most suitable parameters from figures \ref{fig:5} (b) and (c) and explore the variation as a function of  $\Delta/\gamma$. In Fig. \ref{fig:6}(a) and (b) $\bar n$ (in the left y axis) and $\log_{e}g^{(n)}(0)$ (in the right y axis) vs. $\Delta/\gamma$ are plotted. In Fig. \ref{fig:6}(a) a strong two-photon blockade regime at $\Delta/\gamma=\frac{E^{2-'}_{dr}}{2}$ and for $\eta=0.63\gamma$ characterised by $g^{(2)}(0)>1$ and $g^{(3)}(0)<1$ is found. In this regime $g^{(2)}(0)\approx 1$ and $g^{(3)}(0)\approx 0.17$ which indicates an enhanced two-photon blockade. A moderately strong two photon blockade is also found at $\Delta/\gamma=\frac{E^{20'}_{dr}}{2} (g^{(3)}(0)\approx 0.7)$ and $\frac{E^{2+'}_{dr}}{2} (g^{(3)}(0)\approx 0.36)$. In Fig. \ref{fig:6}(b) we show single photon blockade characterised by $g^{(3)}(0)<g^{(2)}(0)<1$, observed at $\Delta/\gamma=\frac{E^{2-'}_{dr}}{2}$ and for $\eta=1.53\gamma$. For the single-photon blockade, $g^{(2)}(0)\approx 0.7$ and  $g^{(3)}(0)\approx 0.31$.

The foregoing results can also be confirmed by comparing the photon-number distributions with the Poisson distribution. In order to verify the two-photon blockade, we plot $\frac{P_n-{\Pi}_n}{\Pi_n}$ with respect to photon number in Fig. \ref{fig:6}(c) which shows that only two-photon probabilities are enhanced ($P_2>\Pi_2$) and all the higher order photon probabilities are suppressed ($P_n<\Pi_n$ for $n>2$). To verify single-photon blockade in (b) we plot the same as (c) in Fig. \ref{fig:6}(d) and notice that only single-photon probabilities are enhanced ($P_1>\Pi_1$) and all the higher order photon probabilities are suppressed ($P_n<\Pi_n$ for $n>1$). It is to be noted that in both the cases (a) and (b), we plot $\bar n$ to confirm the presence of appreciable photon number at the respective points. The rest of the parameters are same as in Fig. \ref{fig:5}. 

\begin{figure}
 \centering
   \includegraphics[width=3.3in]{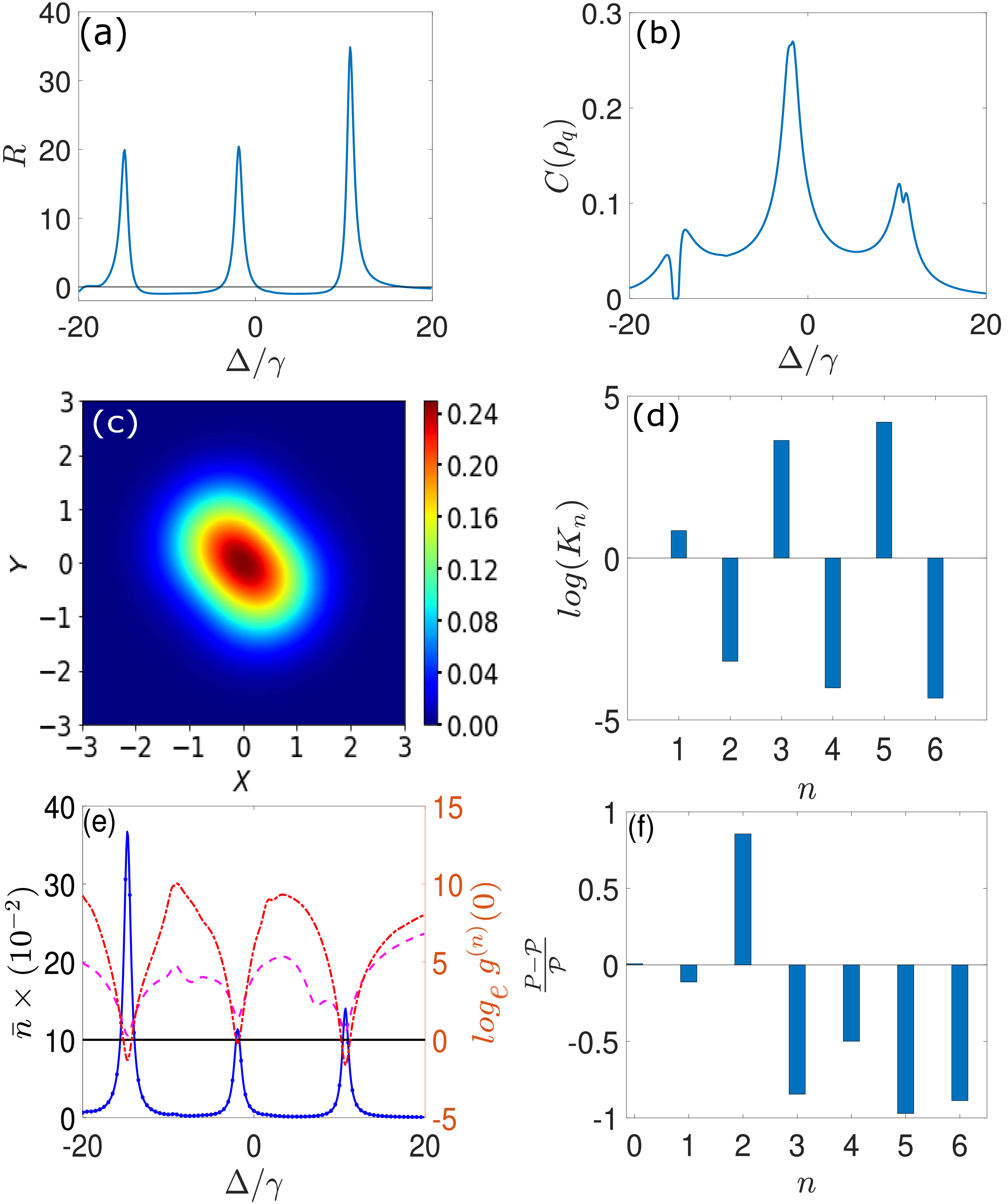} \hspace{0.9em}%
    \caption{(a) $R$ and (b) $C(\rho_q)$ are plotted as a function of $\Delta/\gamma$. (c) Wigner function is plotted in $X$ and $Y$ space, (d) logarithm of the Kleyshko's parameter $K_n$ is plotted as a function of $n$. (e) $\bar{n}$ (solid), $\rm{log}_{e}g^{(n)}(0)$ ($n=2$ $(\rm{dashed})$, $3$  $(\rm{dashed-dotted})$) vs. $\Delta/\gamma$ and (f) $\frac{P_n-\Pi_n}{\Pi_n}$ vs. $n$ are shown. For all these figures $\eta=0.4\gamma$. For (c), (d) and (f)  $\Delta/\gamma=\frac{E^{2-'}_{dr}}{2}$.  The rest of the parameters are same as in Fig. \ref{fig:4}.}
    \label{fig:7}
\end{figure}

Now, to further corroborate our results, we have judiciously chosen a parameter regime and explored all the statistical properties. In Fig. \ref{fig:7}(a) and (b), we have plotted $R$ and $C(\rho_q)$ as a function of $\Delta/\gamma$ and $\eta/\gamma$. We show that at $\Delta/\gamma=\frac{E^{2-'}_{dr}}{2}$ the highest peak in the $R$ vs. $\Delta/\gamma$ curve has a finite concurrence. We also investigate squeezing in the same regime. The Wigner distribution in Fig. \ref{fig:7}(c) shows eliptical profile in the X-Y space in the diagonal direction that indicates the emitted radiation is squeezed. Squeezing is indeed verified by the Kleysko's parameter $K_n$ in Fig. \ref{fig:7}(d) which shows prominent even- odd oscillations in the photon number space. Lastly, we plot $\rm{log}_{e}g^{(n)}(0)$ (right y-axis) and $\bar{n}$ (left y-axis) as a function of $\Delta/\gamma$ in Fig. \ref{fig:7}(e). $\bar{n}$ shows three distinct peaks at $\frac{E^{2-'}_{dr}}{2},\frac{E^{20'}_{dr}}{2},\frac{E^{2+'}_{dr}}{2}$. In these positions $g^{3}(0)<1$ and $g^{2}(0)>1$ indicating a 2P blockade. In these three positions the $g^{2}(0)\approx 1,2.48,1.1$ and the $g^{(3)}(0)$ is achieved as low as $g^{(3)}(0)\approx 0.16, 0.7,0.3$. So, $g^{(2)}(0)$ is one order of magnitude higher than $g^{(3)}(0)$ at $\frac{E^{20'}_{dr}}{2}$ which indicates an enhanced 2P blockade. The strong 2P blockade is again confirmed by plotting $\frac{P_n-\Pi_n}{\Pi_n}$ as a function of photon number. This quantity shows a peak at $n=2$ ($P_2>\Pi_2$) and its values for all higher photon numbers are negative ($P_n<\Pi_n$ for $n>2$) indicating a strong two-photon emission with suppression of higher photon number states.  Thus our system can act as a hyperradiant two-photon squeezed light generator.

\section{Conclusions}\label{conclusions}

In summary, we have studied the quantum properties of a CQED system with a pair of qubits inside a cavity driven by a two-photon drive. We have shown that the system generates strong hyperradiance in the weak-driving regime where finite concurrence between the qubits exists. In the intermediate and strong driving regime, though concurrence may exist, the hyperradiance disappears or becomes very small. So, we  infer that a strong hyperrdiance in the weak driving regime may be associated with two-qubit entanglement, but the existence of entanglement can not necessarily be associated with hyperradiance. 
 We have further demonstrated that the emitted hyperradiant light can be squeezed in quadrature. To confirm squeezing, we have calculated Wigner distribution of the field and Kleyshko's parameter. The Wigner distribution shows eliptical profile while the Kleyshko's parameter exhibits even-odd oscillations as a clear signature of squeezing. We have also explored the influence of  an intra-cavity Kerr-nonlinear medium on the system. The nonlinear medium induces photon blockade that can be employed to control the photon statistics of the field. We have shown that the hyperradiant field can be generated with single as well as two-photon blockades in suitable parameter regimes. The hyperradiant field with two-photon blockade is found to be quadrature-squeezed. Our findings may have  potential applications in quantum communication and quantum networking.

\section*{acknowledgement}
AD is thankful to Subhanka Mal for fruitful scientific discussions.

\appendix
\section{ Analysis using non-hermitian Hamiltonian approach
}\label{analytic}
In what follows, we present approximate analytical expressions for the radiance $R$ of the
transmitted field. Under the scenario of weak driving ($\eta<<
g,\kappa$) the total excitation of the qubit-cavity system
can be assumed to be two as in Refs. \cite{PhysRevA.83.021802,PhysRevA.88.033836}. Under the weak driving limit, the qubit-cavity system remains predominantly in the ground state. So, the terms arising as $\kappa a \rho a^{\dagger}$ and $\gamma \sigma_j\rho\sigma_j^{\dagger}$ in the Lindblad master equation can safely be ignored \cite{PhysRevA.98.013839,PhysRevA.100.062131}.
Under these circumstance, the joint wave function of the pair of qubits and the cavity field can be reasonably approximated in the two-photon
manifold with the ansatz
\begin{equation}
\begin{split}
|\psi_2(t)\rangle &= C_{gg,0}(t)|gg,0\rangle+C_{gg,2}(t)|gg,2\rangle+
 C_{gg,1}(t)|gg,1\rangle+C_{+,1}(t)|+,1\rangle+\\&
 C_{+,0}(t)|+,0\rangle+C_{ee,0}(t)|ee,0\rangle.
\end{split}
\label{psi_2atom}
\end{equation}
and the corresponding Hamiltonian will be
\begin{equation}
\begin{split}
     H^2_{nh} &=(\Delta_c-i\kappa/2)\hat{a}^{\dagger}\hat{a}+(\Delta_a -i\gamma_1/2)\hat{\sigma_1}^{\dagger}\hat{\sigma_1}+(\Delta_a -i\gamma_2/2)\hat{\sigma_2}^{\dagger}\hat{\sigma_2}+\\& g_1(\hat{a}^{\dagger}\hat{\sigma_1}+\hat{a}\hat{\sigma_1}^{\dagger})+g_2(\hat{a}^{\dagger}\hat{\sigma_2}+\hat{a}\hat{\sigma_2}^{\dagger})+
     \chi {\hat{a}}^\dagger{\hat{a}}^\dagger\hat{a}\hat{a}+
     \eta({\hat{a}^{\dagger^2}}+\hat{a}^2).
     \end{split}
     \label{H_2atom}
\end{equation}
The coefficient $C_{lm,n}$ stands for the probability
amplitude of the corresponding state $|lm,n\rangle$.

 We also treat single qubit-cavity system in the two  photon space  with the ansatz wave function
\begin{equation}
\begin{split}
|\psi_1(t)\rangle= C_{g,0}(t)|g,0\rangle+C_{g,1}(t)|g,1\rangle+C_{g,2}(t)|g,2\rangle+
 C_{e,0}(t)|e,0\rangle+C_{e,1}(t)|e,1\rangle
\end{split}
\label{psi_1atom}
\end{equation}
and the non-Hermitian Hamiltonian is given by
\begin{equation}
     H^1_{nh}=(\Delta_c-i\kappa/2)\hat{a}^{\dagger}\hat{a}+(\Delta_a -i\gamma/2)\hat{\sigma_1}^{\dagger}\hat{\sigma_1}+g(\hat{a}^{\dagger}\hat{\sigma_1}+\hat{a}\hat{\sigma_1}^{\dagger})+
     \chi {\hat{a}}^\dagger{\hat{a}}^\dagger\hat{a}\hat{a}+\eta({\hat{a}^{\dagger^2}}+\hat{a}^2).
     \label{H_1atom}
\end{equation}
The coefficient $C_{m,n}$ stands for the probability
amplitude of the corresponding state $|m,n\rangle$.
Time dependent Schroedinger equation incorporating $H^1_{nh}$ and $\psi_1(t)$ yields a set of coupled differential equations as follows
\begin{eqnarray}
i\hbar\dot{C_{g,1}}&=&\Delta C_{g,1}+gC_{e,0}-i\frac{\kappa}{2}C_{g,1}\nonumber\\
i\hbar\dot{C_{g,2}}&=&2\Delta C_{g,2}+\sqrt{2}gC_{e,1}+2\chi C_{g,2}+\sqrt{2}\eta C_{g,0}-i{\kappa}C_{g,2}\nonumber\\
i\hbar\dot{C_{e,0}}&=&\Delta C_{e,0}+g C_{g,1}-i{\gamma}C_{e,0}\nonumber\\
i\hbar\dot{C_{e,1}}&=&2\Delta C_{e,1}+\sqrt{2}gC{g,2}-i(\frac{\kappa}{2}+\frac{\gamma}{2})C_{e,1}.
\label{1_atom}
\end{eqnarray}
In the weak driving limit, we have the relationship $C_{g,0}>>C_{g,2},C_{e,1}>>C_{g,1},C_{e,0}$ since it is a two photon drive. So, we assume that $C_{g,0}\approx 1$. Here we have taken $\Delta_c=\Delta_a=\Delta$.
Again using equations (\ref{H_2atom}) and (\ref{psi_2atom}), we get the coupled differential equations of the form
\begin{eqnarray}
i\hbar\dot{C_{gg,2}}&=&2\Delta C_{gg,2}+2g C_{+,1}+ 2\chi C_{gg,2}+\sqrt{2}\eta C_{gg,0}   -i\kappa C_{gg,2}\nonumber\\
i\hbar\dot{C_{gg,1}}&=&\Delta C_{gg,1}+\sqrt{2}g C_{-,0}+\sqrt{6}\eta C_{gg,3}-\frac{i\kappa}{2} C_{gg,1}\nonumber\\
i\hbar\dot{C_{+,1}}&=&2\Delta C_{+,1}+\sqrt{2}g C_{ee,0}+2g C_{gg,2}+2\chi C_{+,2}+\sqrt{2}\eta C_{+,0}-i(\frac{\kappa}{2}+\frac{\gamma}{2})C_{+,1}\nonumber\\
i\hbar\dot{C_{+,0}}&=&\Delta C_{+,0}+\sqrt{2}g C_{gg,1}-i\frac{\gamma}{2} C_{+,0}\nonumber\\
i\hbar\dot{C_{ee,0}}&=&2\Delta C_{ee,0}+\sqrt{2}g C_{+,1}-i\gamma C_{ee,0}
\label{2_atom}
\end{eqnarray}
 Again, considering weak driving we assume that $C_{gg,0}\approx 1$. We solve these dynamical equations in the steady-state.  The solution of equation (\ref{1_atom}) and (\ref{2_atom}) are given by,
\begin{eqnarray}
C_{g,2}&=&\frac{-\sqrt{2}\eta}{2((\Delta_k+\chi)(\Delta_k+\Delta_\gamma))-g^2}\nonumber\\
C_{e,1}&=&\frac{\eta g}{2((\Delta_k+\chi)(\Delta_k+\Delta_\gamma))-g^2}
\end{eqnarray}
\begin{eqnarray}
C_{gg,2}&=&\frac{-\sqrt{2}\eta(\Delta_l^2+\Delta_k\Delta_l-g^2)}{2(\Delta_l(\Delta_k+\chi)(\Delta_k+\Delta_l))-g^2(\Delta_k+2\Delta_l+\chi)}\nonumber\\
C_{+,1}&=&\frac{-\sqrt{2}\Delta_l\eta g}{2(\Delta_l(\Delta_k+\chi)(\Delta_k+\Delta_l))-g^2(\Delta_k+2\Delta_l+\chi)}
\label{coefficients}
\end{eqnarray}
For the sake of simplicity, we have defined two effective complex detunings as $\Delta_{\kappa}=\Delta-\frac{i\kappa}{2}$ and $\Delta_{\gamma}=\Delta-\frac{i\gamma}{2}$. Now, using this solutions we obtain the expression of radiance as follows,
\begin{equation}
\begin{split}
R &=\left(\frac{-8\eta^2(4g^2+16\Delta^2+(\kappa+\gamma)^2)}{|A+iB|^2}+ \frac{4\eta^2(|C|^2+8g^2|(2\Delta-i\gamma)|^2)}{|A'+iB'|^2}\right)\times \\& \frac{D}{8\eta^2(4g^2+16\Delta^2+(\kappa+\gamma)^2)}
\end{split}
\label{radiance}
\end{equation}
where 
\begin{eqnarray}
A&=&\kappa\gamma+4g^2+\kappa^2-8\Delta\chi-8\Delta^2\nonumber\\
B&=&2(\kappa\chi+\Delta\gamma+3\Delta\kappa+\chi \gamma)\nonumber\\
A'&=&-16\Delta^3-16\chi\Delta^2+24 \Delta g^2+2\Delta\kappa^2+8\Delta\kappa\gamma+2\Delta\gamma^2+8\chi g^2+2\chi\kappa\gamma+2\chi\gamma^2\nonumber\\
B'&=&12\Delta^2\kappa+12\Delta^2\gamma+4\chi\Delta\kappa+12\chi\Delta\gamma-4g^2\kappa-8g^2\gamma-\kappa^2\gamma-\kappa\gamma^2\nonumber\\
C&=&(8\Delta^2-6i\Delta \gamma-2i\Delta \kappa-4g^2-\gamma^2-\kappa\gamma)\nonumber\\
D&=&(4(\kappa+\gamma)^2(\Delta+\chi)^2+(4g^2+\kappa^2)^2)+64\Delta(\Delta^3+2\Delta^2\chi-\Delta g^2+\Delta \chi^2-g^2 \chi)+ E\nonumber\\ 
E&=&(16\Delta^2 \kappa^2+8 g^2 \kappa \gamma +2 \kappa^3 \gamma +\kappa^2 \gamma^2)\nonumber\\
\label{R_constants}
\end{eqnarray}
 \begin{figure}
    \centering 
    \includegraphics[width=0.3\linewidth]
    {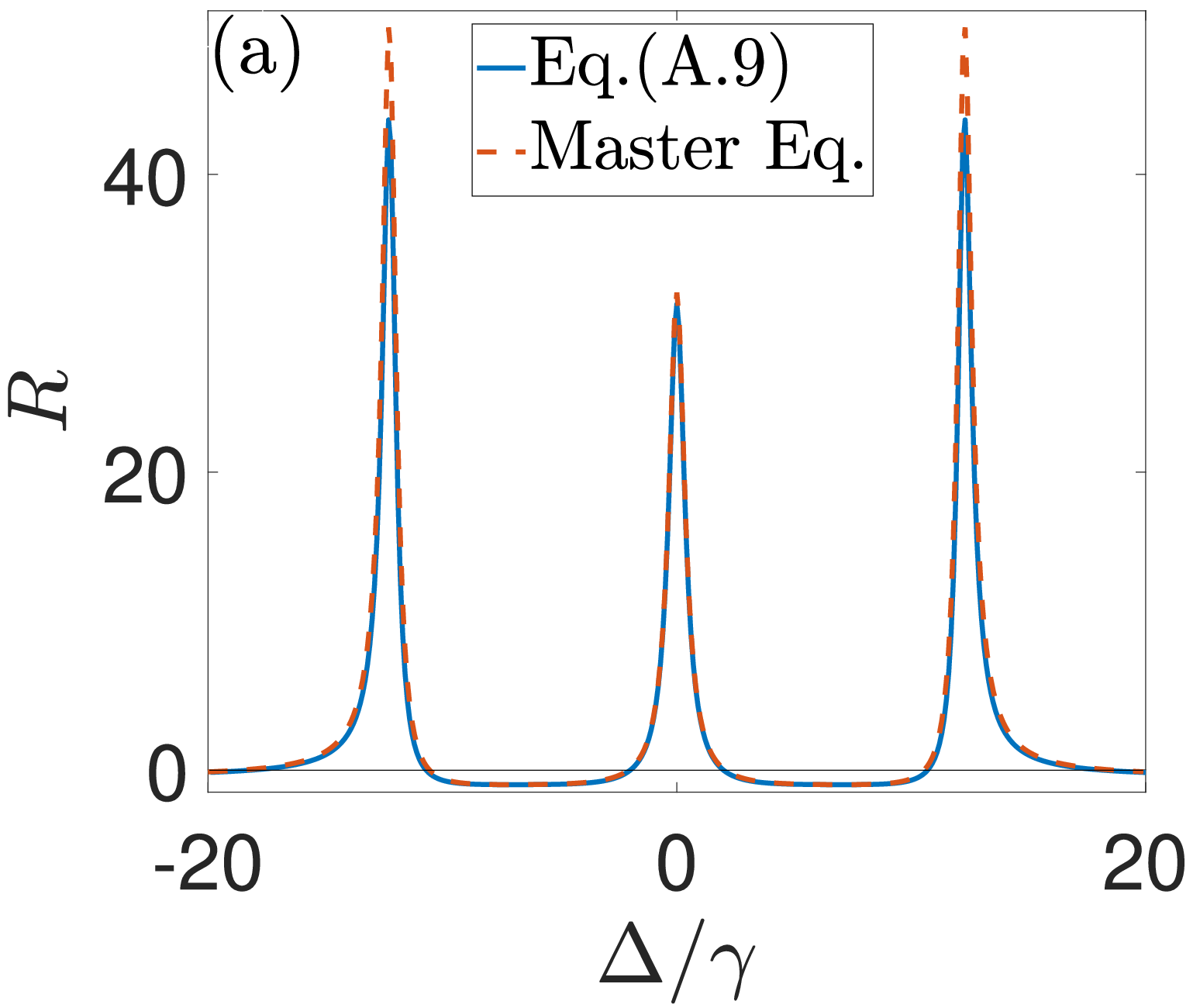} 
    \includegraphics[width=0.32\linewidth]
    {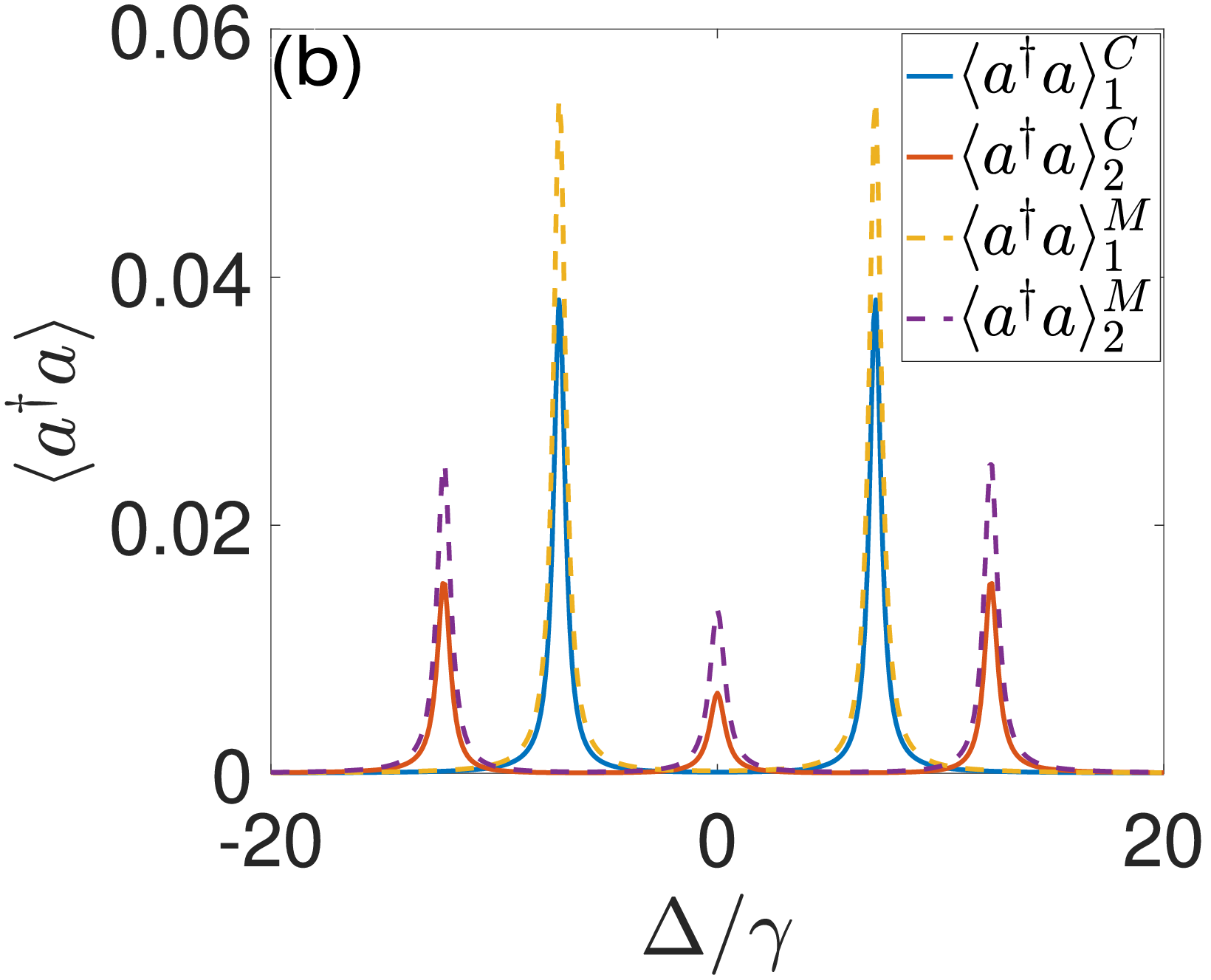}
    \includegraphics[width=0.32\linewidth]
    {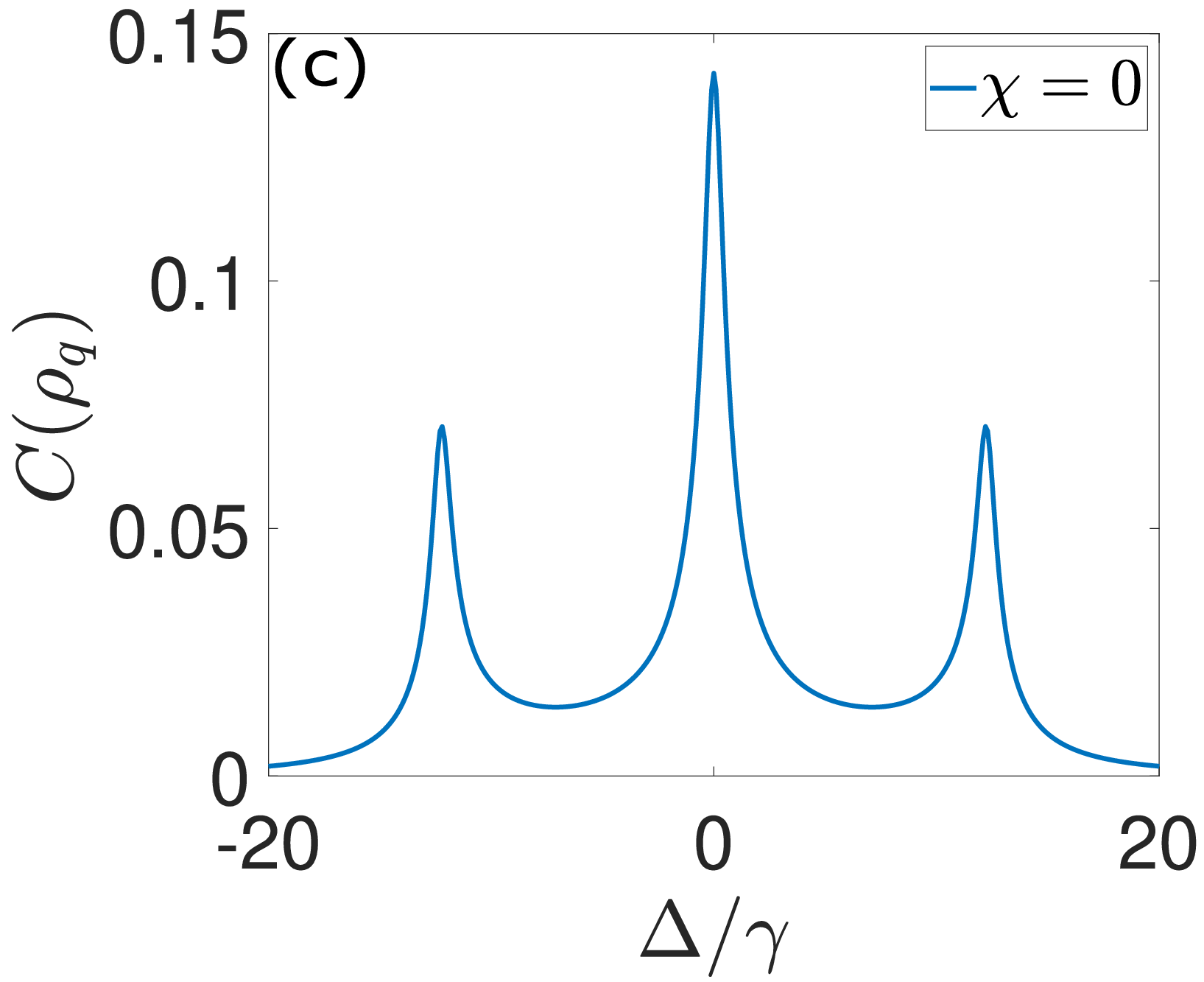}
    \caption{(a) $R$ and (b) ${{\langle a^{\dagger}a\rangle}_i}^{M(C)}$ ($i=1,2$) are plotted as a function of $\Delta/\gamma$ by solving both equations (\ref{radiance}) (solid) and the master equation (\ref{eq1}) (dashed). Here $M(C)$ stands for master equation (complex hamiltonian) and $i$ refers to the number of qubits present in the cavity. (c) Concurrence $C(\rho_q)$ is plotted as a function of $\Delta/\gamma$. For (a), (b) and (c) $\eta/\gamma=0.1$ and the rest of the parameters are same as in Fig. \ref{fig:1}.}
    \label{fig:analytic_ki0}
\end{figure}
Figures. \ref{fig:analytic_ki0}(a) and \ref{fig:analytic_ki0}(b) show $R$ and $\langle a^{\dagger}a\rangle$ as a function of $\Delta/\gamma$ at $\eta=0.4$ using both perturbative analytical result given by equation (\ref{radiance}) and numerically solving master  equation (\ref{eq1}). As shown in these figures, these two results match quite well except slight mismatch in the amplitude or intensity. This is because the states included in the complex hamiltonian approach excludes the gain pathways unlike the master equation approach of equation (\ref{eq1}).
 Figure \ref{fig:1}(b) can be understood from Fig. \ref{fig:analytic_ki0}(c) where two-qubit concurrence $C(\rho_q)$ is plotted as a function of $\Delta/\gamma$ for $\eta=0.1\gamma$. Three peaks appearing at the transition frequencies as shown in Fig. \ref{fig:analytic_ki0}(a) correspond to moderate concurrence. The concurrence goes down when the system is not in the hyperradiant regime or in subradiant regime. Therefore, it is worth noting that the radiance and concurrence seems to have a one-to-one connection as far as the low driving regime is concerned.
\begin{figure}
    \centering 
    \includegraphics[width=0.3\linewidth]
    {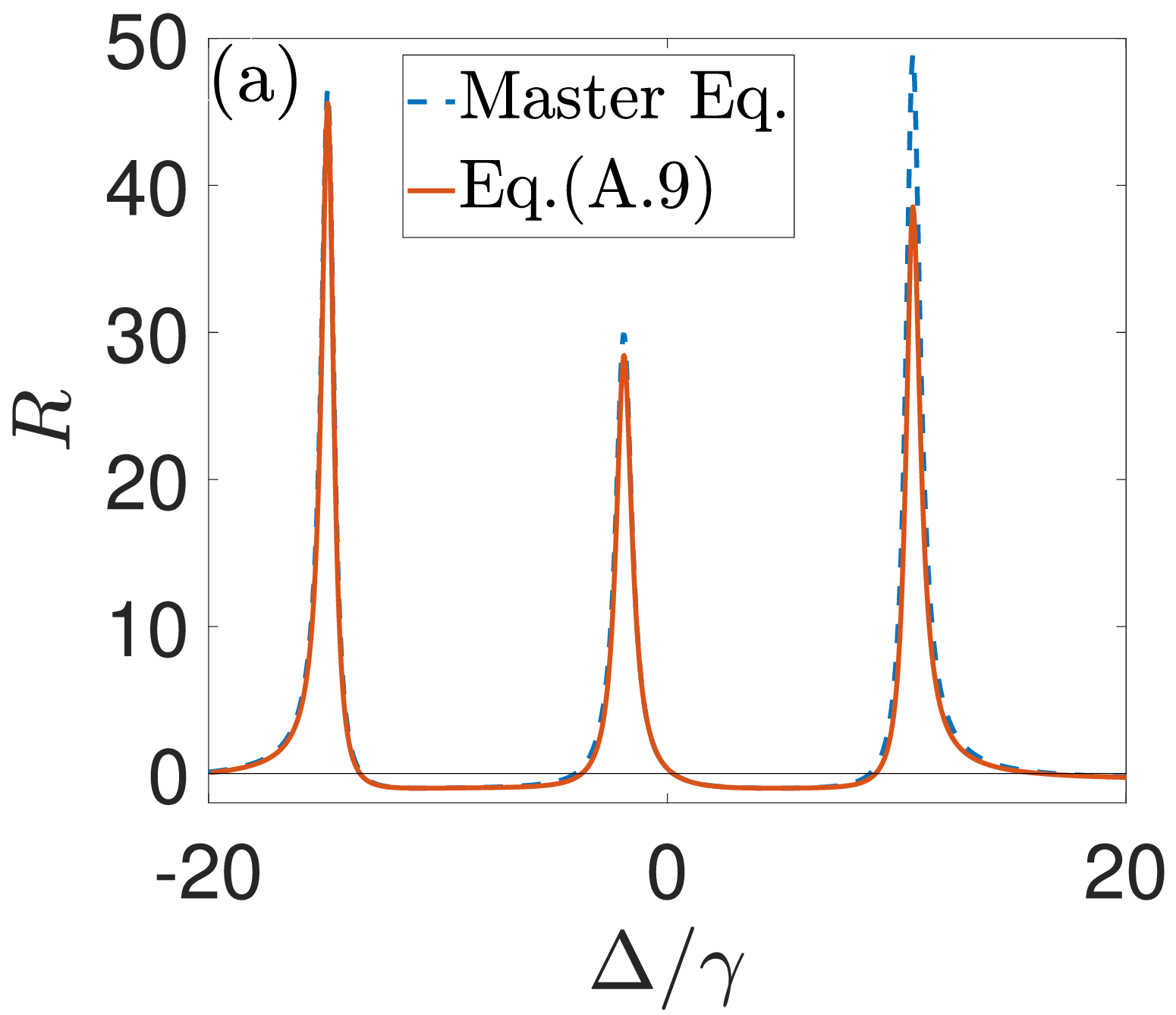} 
    \includegraphics[width=0.32\linewidth]
    {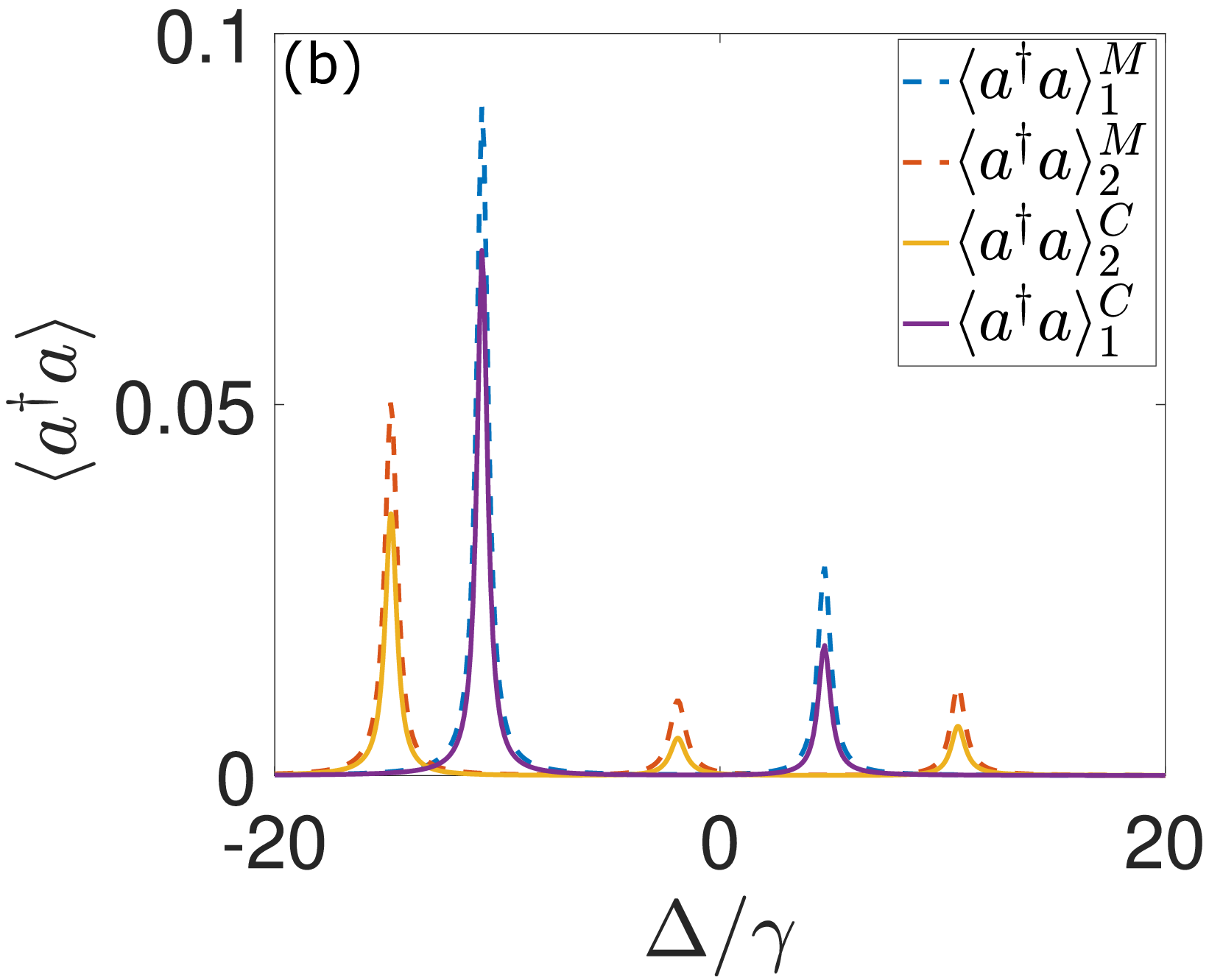}
    \includegraphics[width=0.32\linewidth]
    {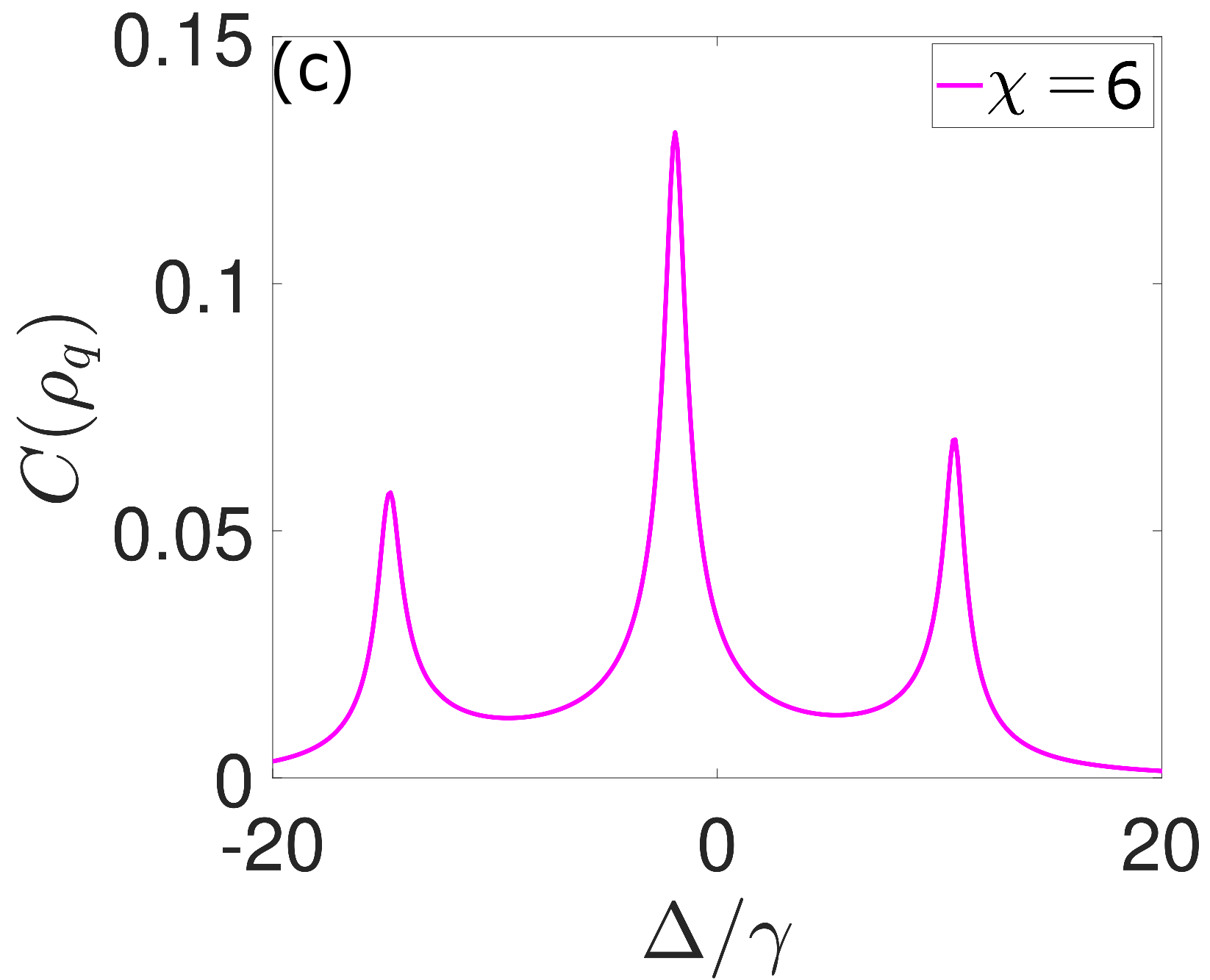}
    \caption{(a) $R$ and (b) ${{\langle a^{\dagger}a\rangle}_i}^{M(C)}$ $(i=1,2)$ are plotted as a function of $\Delta/\gamma$ by solving both equations (\ref{radiance}) (solid) and the master equation (\ref{eq1}) (dashed). Here $M(C)$ denotes master equation (complex hamiltonian) and $i$ refers to the number of qubits present in the cavity. (c) Concurrence ($C(\rho_q)$) is plotted as a function of $\Delta/\gamma$.  For (a), (b) and (c) $\eta/\gamma=0.1$ and the rest of the parameters are same as in fig. \ref{fig:4}.}
    \label{fig:analytic}
\end{figure}
To confirm the hyperradiance behaviour in the presence of a Kerr medium, we have plotted $R$ in Fig. \ref{fig:analytic}(a) as well as photon populations in \ref{fig:analytic}(b) as a function of $\Delta/\gamma$. Both perturbative and master equation based results are in good agreement except slight mismatch in the amplitudes as discussed in sec. \ref{discussion}(\ref{chi0}). We notice that the incorporation of nonlinearity $\chi$ has enhanced the radiance and the population. Next, We have plotted the concurrence $C(\rho_q$) as a function of $\Delta/\gamma$ for $\eta=0.1$ in Fig. \ref{fig:analytic}(c). This figure justifies the fact that for detunings that correspond to the resonant transitions between the dressed states, the entanglement is quite significant. 
\begin{figure}
 \centering
   \includegraphics[width=3in]{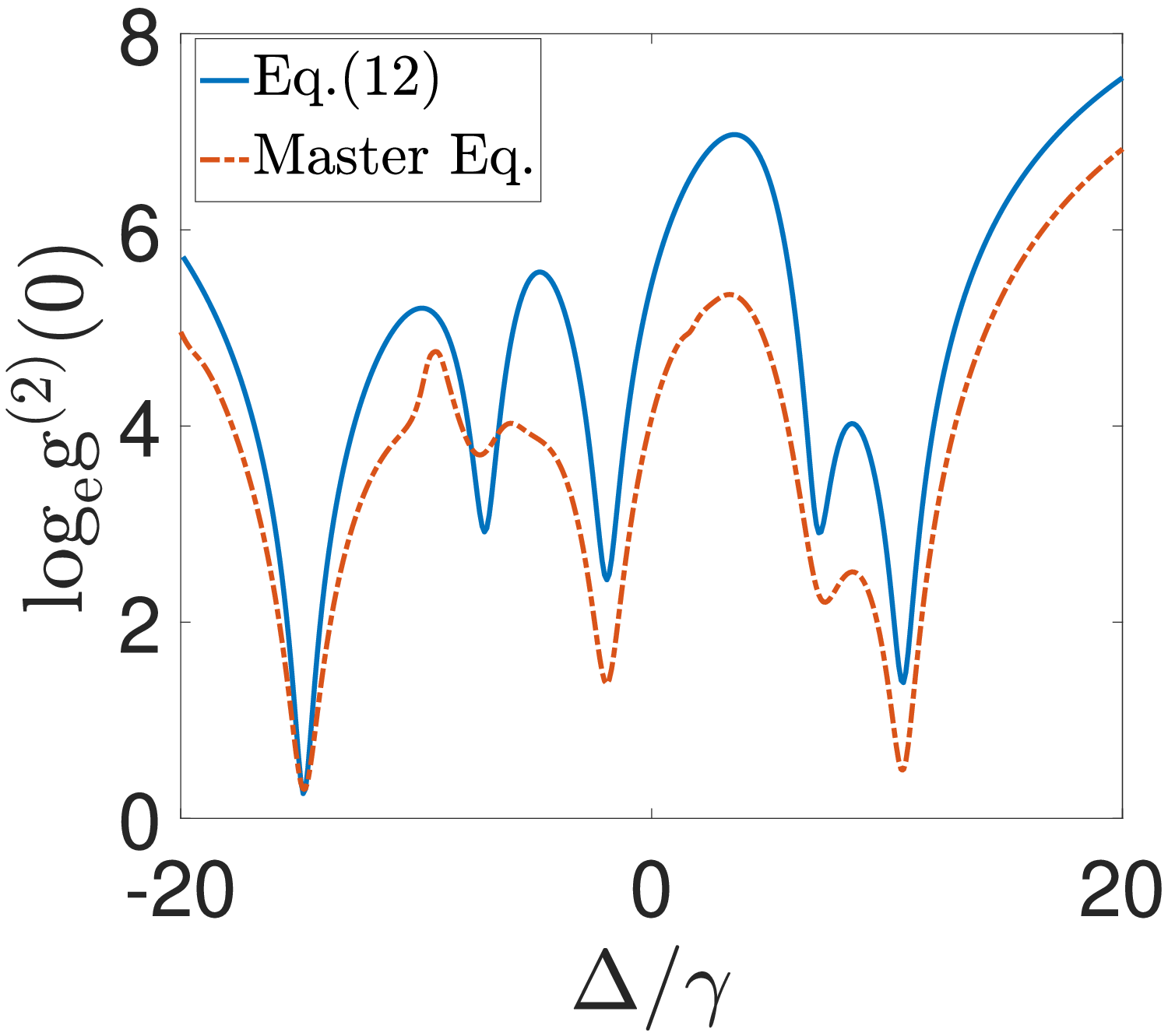} \hspace{0.9em}%
    \caption{(a) $\rm{log}_{e}g^{(2)}(0)$ is plotted as a function of $\Delta/\gamma$ for $\eta=0.4\gamma$ by solving both equations (\ref{g20_C}) (solid) and the Master equation (\ref{eq1}) (dashed). The rest of the parameters are same as in Fig. \ref{fig:4}.}
    \label{fig:analytic_g20}
\end{figure}
To further corroborate the numerical results we have plotted $\rm{log}_{e}g^{(2)}(0)$ as a function of $\Delta/\gamma$ using  both the equations (\ref{g20_C}) and (\ref{eq1}) in Fig. \ref{fig:analytic_g20} for the same parameter regime as in Fig. \ref{fig:7}. Both the results are in good agreement except slight mismatch in the amplitudes for the same reason as mentioned earlier.

\bibliographystyle{iopart-num.bst}
\bibliography{cite2}
\end{document}